\documentclass[aps,prl,twocolumn,amsfonts,amssymb,amsmath,tightenlines,a4paper,notitlepage,groupedaddress,superscriptaddress]{revtex4-1}
\pdfoutput=1
\usepackage{graphicx}
\usepackage{hyperref}
\usepackage[utf8]{inputenc}
\usepackage{ulem}
\usepackage{bm}

\newcommand{\dd}[0]{\mathrm{d}}

\usepackage{color}

\definecolor{ccomments}{rgb}{1,0.2,0.2}

\definecolor{cmissing}{rgb}{1,0.5,0.3}

\usepackage{epstopdf}

\definecolor{cream}{RGB}{222,217,201}

\usepackage{color}

\definecolor{ccomments}{rgb}{1,0.2,0.2}

\definecolor{cmissing}{rgb}{1,0.5,0.3}

\begin{document}


\title{Bridging microscopic cell dynamics to nematohydrodynamics of cell monolayers}
    \author{Aleksandra Arda\v{s}eva}
	\affiliation{The Niels Bohr Institute, University of Copenhagen, Copenhagen, Denmark}
	
	\author{Romain Mueller}
	\affiliation{Rudolf Peierls Centre for Theoretical Physics, University of Oxford, UK}
	
	\author{Amin Doostmohammadi}
	\email{doostmohammadi@nbi.ku.dk}
	\affiliation{The Niels Bohr Institute, University of Copenhagen, Copenhagen, Denmark}

\begin{abstract}
    It is increasingly being realized that liquid-crystalline features can play an important role in the properties and dynamics of cell monolayers. Here, we present a cell-based model of cell layers, based on the phase-field formulation, that connects mechanical properties at the single cell level to large-scale nematic and hydrodynamic properties of the tissue. In particular, we present a minimal formulation that reproduces the well-known bend and splay hydrodynamic instabilities of the continuum nemato-hydrodynamic formulation of active matter, together with an analytical description of the instability threshold in terms of activity and elasticity of the cells. Furthermore, we provide a quantitative characterisation and comparison of flows and topological defects for extensile and contractile stress generation mechanisms, and demonstrate activity-induced heterogeneity and spontaneous formation of gaps within a confluent monolayer. Together, these results contribute to bridging the gap between micro-scale cell dynamics and tissue-scale collective cellular organisation.
    
\end{abstract}

\maketitle

\section{Introduction}
Collective cell migration is crucial for a wide variety of biological scenarios, ranging from development \cite{heisenberg2013forces} to wound healing \cite{brugues2014forces} and pathological conditions, such as cancer metastasis \cite{friedl2012classifying}. The profound role of mechanical forces in regulating such behaviour is being increasingly recognized in a variety of cell types \cite{ladoux2017mechanobiology}. On the other hand, these collective behaviours are to a large degree generic and can be described by different classes of theoretical models that target specific levels of description. In particular, there is a well-documented connection between liquid crystal theories and the description of epithelial cells at large scales \cite{doostmohammadi2021physics, duclos2018spontaneous, kawaguchi2017topological,prost2015active,doostmohammadi2018active}.

Nematic liquid crystals consist of rod-like particles, which under certain conditions align in a given direction. This gives rise to quasi long-range nematic order while singularities in orientational alignment lead to topological defects. Theories of nematic liquid crystals have recently found application in different biological systems. For example, the local orientation of the actin fibres in the tissue can be effectively described as a nematic liquid crystal \cite{maroudas2020topological}, and elongated fibroblasts at high densities show nematic alignment and topological defects due to steric interactions \cite{duclos2018spontaneous}. Even though epithelial cells have no obvious microscopic nematic degree of freedom, several flow features of cell layers have been successfully reproduced in continuum theories of active liquid crystals. This includes flow fields around cell division events~\cite{doostmohammadi2015celebrating}, long-range vortex lattices in dividing cell layers~\cite{rossen2014long, doostmohammadi2016stabilization}, chaotic flow features within the active turbulent state~\cite{blanch2018turbulent, lin2021energetics}, and formation of topological defects, that have been linked with death and extrusion events in monolayers of MDCK cells \cite{saw2017topological}. However, there is still little understanding on how emergent flows relate to microscopic dynamics of cells. 

The mechanisms by which hydrodynamic modes emerge from overdamped microscopic dynamics is well understood for self-propelled point particles with aligning interactions, and there are now powerful methods that allow consideration of a large class of such models \cite{marchetti2013hydrodynamics, peshkov2014boltzmann}. There have been efforts to apply similar ideas to cellular layers \cite{lee2011crawling, lee2011advent, gov2009traction} but how these models relate to a hydrodynamic description is less clear because dense cellular systems are far from the dilute limit and steric interactions are important. More complex physical descriptions of dense layers have also been explored \cite{camley2017physical} but most of the investigations concentrated around rigidity transition and jamming \cite{bi2016motility, henkes2011active} or tissue reshaping by the extracellular matrix \cite{etournay2015interplay}. 

Lately, the phase-field formalism has proven successful in reproducing various experimental phenomenologies, including mechanical stress patterns \cite{saw2017topological}, as well as the cell-substrate and cell-cell interaction forces in epithelial cell layers \cite{balasubramaniam2021investigating}. In these methods, the cells are modelled as active deformable droplets in two-dimensions whose boundaries are defined using a phase-field potential which allows for the precise description of the intercellular interfaces and forces. Compared to models where the individual cells are described using a tesselation of the plane \cite{bi2016motility, barton2017active}, it has the advantages that it can accommodate large shape deformations of the cells independently of the location of their center-of-masses and that it naturally allows for the description of tissue boundaries, including gaps within the tissue.

The phase-field approach has been used to study the behaviour of cellular systems ranging from single cells \cite{aranson2016physical} to multicellular structures \cite{najem2016phase} and cell monolayers \cite{basan2013alignment, sepulveda2013collective, monfared2021mechanics, mueller2019emergence, zhang2020active}. However, most of these models introduce polar driving of the individual cells as a source of activity and/or polar interactions between cells, and it is unclear if such microscopic dynamics can be properly modelled by a hydrodynamic continuum model of nematic liquid crystals at macroscopic scales. A formulation based on coupling active stresses to cell shape deformations showed the possibility of capturing some aspects of topological defect formation in isotropic cell layers~\cite{mueller2019emergence}. Recently, this approach has been extended to account for both extensile and contractile stresses, noting how these stresses contribute to cell shape deformations~\cite{zhang2021active}. However, the mechanisms of topological defect formation and the link with the well-established instabilities in continuum formulations of active nematics remain unresolved.

In this paper, we present a simple microscopic model of a dense 2D cellular layer that reproduces the phenomenology of hydrodynamic theories of active nematic liquid crystals and introduce a generic model of force transmission at the cellular interfaces, which allows us to connect tissue-level quantities to the microscopic description. We then use insights from the continuum theory of active nematic liquid crystals to introduce a minimal form of local dipolar interaction between cells and show that our model develops the correct hydrodynamic behaviour at macroscopic scales. In particular, we show that this formulation reproduces the typical bend and splay instabilities of active nematic liquid crystals theories and that it produces topological defects and associated flow features that are similar to the ones observed experimentally. Furthermore, within a certain parameter regime, we show that the competition between active stress generation and the elasticity of individual cells results in the spontaneous formation of gaps within the confluent layers.

The paper is organized as follows: In Section~\ref{sec:meth} we describe the details of the model and the minimal description of the cell-generated forces and intercellular interactions. In Section~\ref{sec:bend} we present numerical simulations of the system that reproduce fundamental hydrodynamic instabilities of continuum active nematics, followed by a linear stability analysis in Section~\ref{sec:LSA} that sheds light on the mechanism of the generation of such hydrodynamic modes within our cell-based description. In Section~\ref{sec:def} we employ numerical simulations to move beyond these instabilities and analyse more closely the dynamics of topological defects, followed by characterization of the flow features of active turbulence in the cell-based model in Section~\ref{sec:flow}. Finally, we present quantitative results of gap formation in extensile and contractile systems in Section~\ref{sec:gap}, followed by concluding remarks and discussions in Section~\ref{sec:conc}.

\section{The phase-field model\label{sec:meth}}
Following Mueller et al.~\cite{mueller2019emergence}, we represent each cell in an epithelium monolayer via a separate phase field, $\phi_i$. The dynamics of each phase field is governed by
\begin{equation}
\label{eq:dynamics}
\partial_t \phi_i + \vec{v}_i \cdot \vec{\nabla} \phi_i = - \frac{\delta \mathcal{F}}{\delta \phi_i}, \qquad i=1, \ldots, N,
\end{equation}
where $N$ is the total number of cells and $\vec{v}_i$ is the total velocity of the cell $i$. The free energy, $\mathcal{F}$, governs the dynamics of the cell interfaces and has three contributions:
\begin{equation}\label{eq:F}
    \mathcal{F} = \mathcal{F}_{\text{CH}} + \mathcal{F}_{\text{area}} + \mathcal{F}_{\text{rep.}}.
\end{equation}
The first term in eqn \eqref{eq:F} is the Cahn-Hilliard free energy, which stabilizes the interface between exterior and interior of each cell. It is defined as 
\begin{equation}\label{eq:FCH}
    \mathcal{F}_{\text{CH}} = \sum_i \frac \gamma \lambda \int \textup{d}\vec{x}  \left\{ 4 \phi_i^2 (1-\phi_i)^2 + \lambda^2 (\vec{\nabla} \phi_i)^2 \right\},
\end{equation}
where $\lambda$ is the interface width and $\gamma$ is the surface tension that characterizes the cell elasticity~\cite{palmieri2015multiple}. The double-well potential in the first term of eqn \eqref{eq:FCH} makes $\phi_i = 1$ for regions inside the cells and $\phi_i = 0$ for the outside regions. The cell boundary occurs at $\phi_i = 0.5$. This formulation is guided by simplicity and models the cell boundary as a diffuse interface, but could be easily extended to model sharp interfaces~\cite{PMID:4273690, shao2010computational, biben2003tumbling}.

The second term in eqn \eqref{eq:F}, $\mathcal{F}_{\text{area}}$, defines a soft area constraint for each cell and is given by
\begin{equation}\label{eq:Farea}
    \mathcal{F}_{\text{area}} = \sum_i \mu \Big( 1 - \frac{1}{\pi R^2}\int \textup{d}\vec{x}\, \phi_i^2 \Big)^2,
\end{equation}
where $\mu$ controls the relaxation of area changes and $R$ is the cell radius. $\mathcal{F}_{\text{area}}$ ensures that cell areas, $A_i=\int \textup{d}\vec{x} \phi_i^2$, are close to circle, i.e. $\pi R^2$. Note that area conservation is not implemented at the level of eqn~\eqref{eq:dynamics}, i.e. even though cells are mostly incompressible, the cell area can be dramatically altered, for example, squeezed by neighbouring cells or stretched on the substrate~\cite{peyret2019sustained}.

Finally, the repulsion term, $\mathcal{F}_{\text{rep.}}$, penalizes regions where two cells overlap:
\begin{equation}
    \mathcal{F}_{\text{rep.}} = \sum_i \sum_{j \neq i} \frac{\kappa}{\lambda} \int \textup{d} \vec{x} \, \phi_i^2 \phi_j^2\,.
\end{equation}
Here, $\kappa$ denotes the strength of the repulsion. Normalisation has been chosen such that the width of the interfaces at equilibrium is given by $\lambda$ and that the properties of the cells are roughly preserved when $\lambda$ is rescaled.

\begin{figure}[h]
    \includegraphics[width=\linewidth]{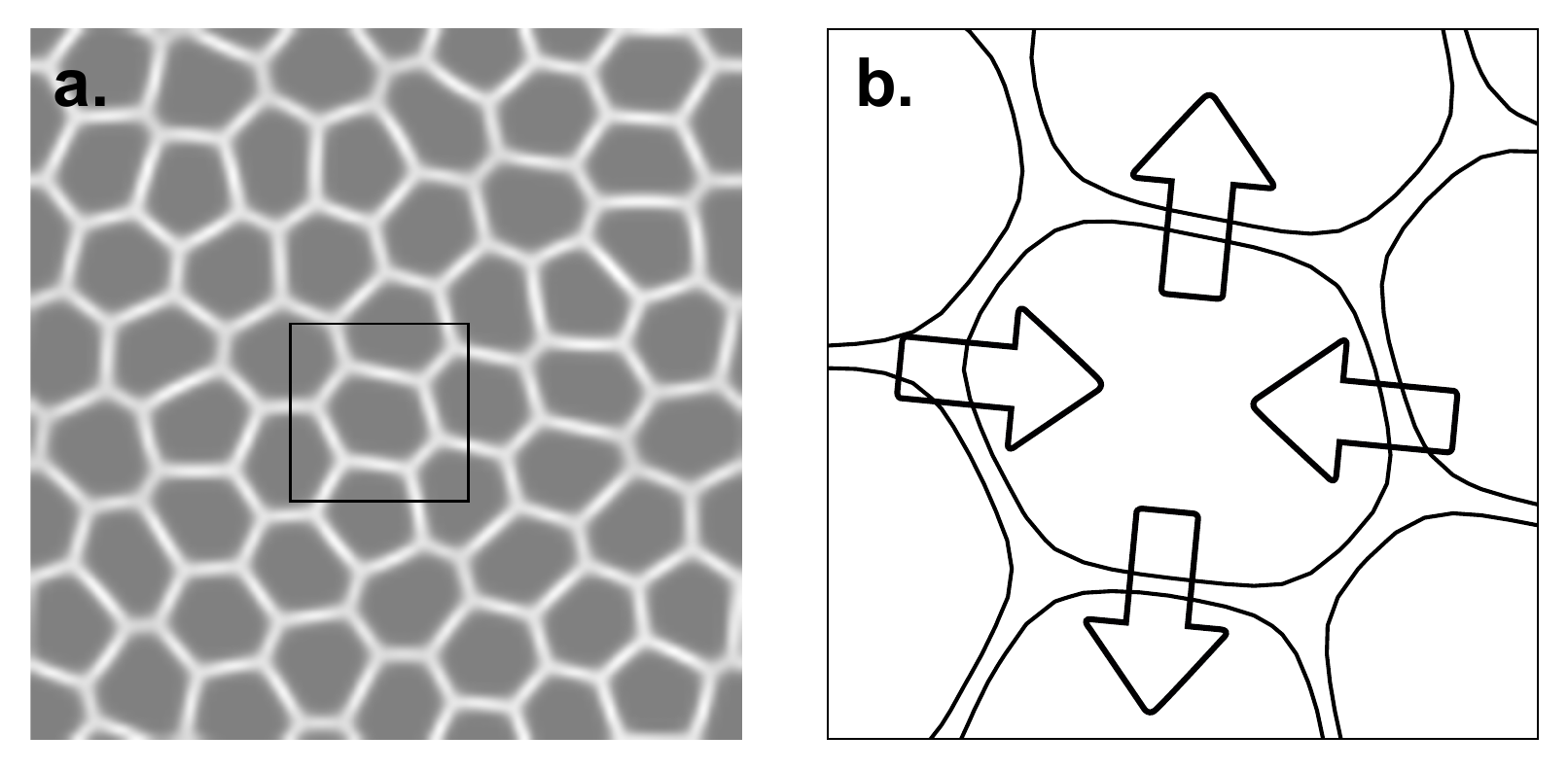}
    \caption{\label{fig:interfaces}%
        (a) Interfaces between cells defined as the overlap $\sum_{i\neq j}\phi_i \phi_j$ and (b) contours $\phi_i=1/2$ with an illustration of the contractile force exerted by the cell in the centre on its neighbours.
    }
\end{figure}

The above mentioned free energies describe the cellular interfaces, as shown in Fig.~\ref{fig:interfaces}a, and specify the equilibrium properties of the interfaces including surface tension and interface width. To drive the system away from the equilibrium we define the cell velocities $\vec{v}_i$ in eqn~\eqref{eq:dynamics} based on the force balance for an overdamped system:
\begin{equation}
\label{eq:force balance}
\xi \vec{v}_i = \vec{F}^{\text{\text{int.}}}_i,
\end{equation}
where $\xi$ is the substrate friction coefficient and $\vec{F}^{\text{\text{int.}}}_i$ is the total force acting on the interface of the cell $i$. 
When modelling epithelia from a material science perspective, these microscopic interface forces are usually defined in terms of a macroscopic tissue stress tensor, $\bm{\sigma}_{\text{tissue}}$.
In our phase-field formulation, this corresponds to
\begin{equation}
\label{eq:Ftot}
\vec{F}^{\text{\text{int.}}}_i = \int \dd \vec{x}\, \phi_i \; \vec{\nabla} \cdot \boldsymbol{\sigma}_{\text{tissue}} = - \int \dd \vec{x}\, \boldsymbol{\sigma}_{\text{tissue}} \cdot \vec{\nabla} \phi_i.
\end{equation}
These expressions can be interpreted as follows: the first is the integral of the local force, $\vec{\nabla}\cdot \bm{\sigma}_{\text{tissue}}$, weighted by the phase-field, $\phi_i$, and the second is the integral of the force exerted by the stress tensor on the vector $-\vec{\nabla} \phi_i$ normal to the interface.
The eqn \eqref{eq:Ftot} connects the local forces at the level of the individual cells with the properties of the cell monolayer.
In the limit of sharp interfaces, the integral tends to a contour integral over the cell boundary and leads to the usual expressions for the local force in terms of the stress tensor, see Section \ref{sec:LSA} below.

In order to reproduce the phenomenology of active nematic liquid crystals, we include local dipolar interactions into $\bm{\sigma}_{\text{tissue}}$.
To do so, we associate internal nematic tensor, $\bm{Q}_i$, to each cell describing the  orientational alignment of the cell.
The tissue stress tensor is defined as follows,
\begin{equation}
    \label{eq:stress}
    \bm{\sigma}_{\text{tissue}} = - P \bm{I} - \zeta \bm{Q},
\end{equation}
where $\bm{Q} = \sum_i \bm{Q}_i \phi_i$ is tissue's nematic field, $P = \sum_i \partial \mathcal{F}/ \partial\phi_i$ is the pressure field, and $\bm{I}$ is the identity tensor.
The first term in eqn~\eqref{eq:stress} induces steric repulsion between cells, while the second term mirrors the active term found in continuum theories of active liquid crystals~\cite{marchetti2013hydrodynamics}.
It can be interpreted as a dipolar force density distributed along the cells interfaces such that each cell pushes or pulls its neighbours depending on the direction of their contact area with respect to the nematic tensor, as demonstrated schematically in Fig.~\ref{fig:interfaces}b.

We now introduce the dynamics of the internal nematic degree of freedom, $\bm{Q}_i$, of each cell.
In the continuum theory, most of the phenomenology of active liquid crystals depends on a balance between the restoring elastic forces and flow alignment of the nematic tensor.
By writing $\bm{Q}_i = 2 \left(\vec{n}_i \otimes \vec{n}_i - \vec{n}_i^2\, \bm{I} /2\right)$ with $\vec{n}_i = (\cos \theta_i, \sin \theta_i)$, we can mirror these two components in our model and define the following dynamics of the angle, $\theta_i$:
\begin{equation}
\label{eq:theta}
\partial_t \theta_{i} = K \tau_i + J \omega_i,
\end{equation}
where the torques are given by
\begin{align}
    \label{eq:torques}
    \tau_i = \frac 1 \lambda \int \textup{d}\vec{x}\; \bm{Q}_i \phi_i \wedge \bm{Q}, \quad
    \omega_i = \int \textup{d}\vec{x}\; \vec{v} \wedge \vec{\nabla} \phi_i,
\end{align}
and $K$ is the nematic elastic parameter and $J$ is the nematic flow alignment strength.
We have defined the tissue velocity as $\vec{v} = \sum_i \vec{v}_i \phi_i$ and $\bm{A} \wedge \bm{B} = A_{xx} B_{xy} - A_{xy} B_{xx}$ for symmetric traceless matrices $\bm{A}$ and $\bm{B}$, while $\vec{v} \wedge \vec{\nabla} \phi_i$ is the usual wedge product.
The first torque, $\tau_i$, aligns $\bm{Q}_i$ to the tissue nematic tensor, $\bm{Q}$, and induces an elastic restoring force favouring the homogeneous state.
The second torque, $\bm{\sigma}_{\text{tissue}}$, rotates $\bm{Q}_i$ with the local vorticity computed as the integral of the neighbouring cells velocity projected on the interface.
Note that our definitions are local in the sense that each torque is given by an integral along the cell interface with its neighbours.

We simulate eqns~\eqref{eq:dynamics} and~\eqref{eq:theta} using an open-source software \texttt{Celadro} (available on \texttt{GitHub}: \texttt{https://github.com/rhomu/celadro}), which utilizes a finite difference scheme on a square lattice with a predictor-corrector step (for details of the implementation, see Mueller et al. \cite{mueller2019emergence}). For all the simulations presented here, we initialize each cell with a director along the $x$-direction. We simulate the dynamics of 896 cells of radius $R=8$ in a $400\times400$ periodic domain. Cells are placed randomly and relaxed for 200 time steps under passive dynamics to reach confluence. Simulations are run for 15000 time steps with the following parameters/parameter ranges (unless stated otherwise): $\mu = 0.03$, $\lambda =8$, $\kappa = 0.2$,  $\xi = 1$, $K = 2\times10^{-5}$, $J = 4\times 10^{-3}$, $\zeta \in \pm[3\times 10^{-4},8\times10^{-2}]$, $\gamma \in [0.01,0.1]$, chosen within the range that was previously shown to reproduce defect flow fields in epithelial layers~\cite{balasubramaniam2021investigating}. 

\section{Results}
To demonstrate the new features of the developed model and highlight the consistency with the continuum model, we begin by numerically investigating the mechanism and dynamics of the system starting from a quiescent state. 
In our analysis, we explore the behaviour of the model for a range of activity parameter values, $\zeta$, (both positive and negative) and consider different values of the cell elasticity $\gamma$.

\subsection{Bend and splay instabilities\label{sec:bend}}
\begin{figure}[tbh!]
\includegraphics[width=\linewidth]{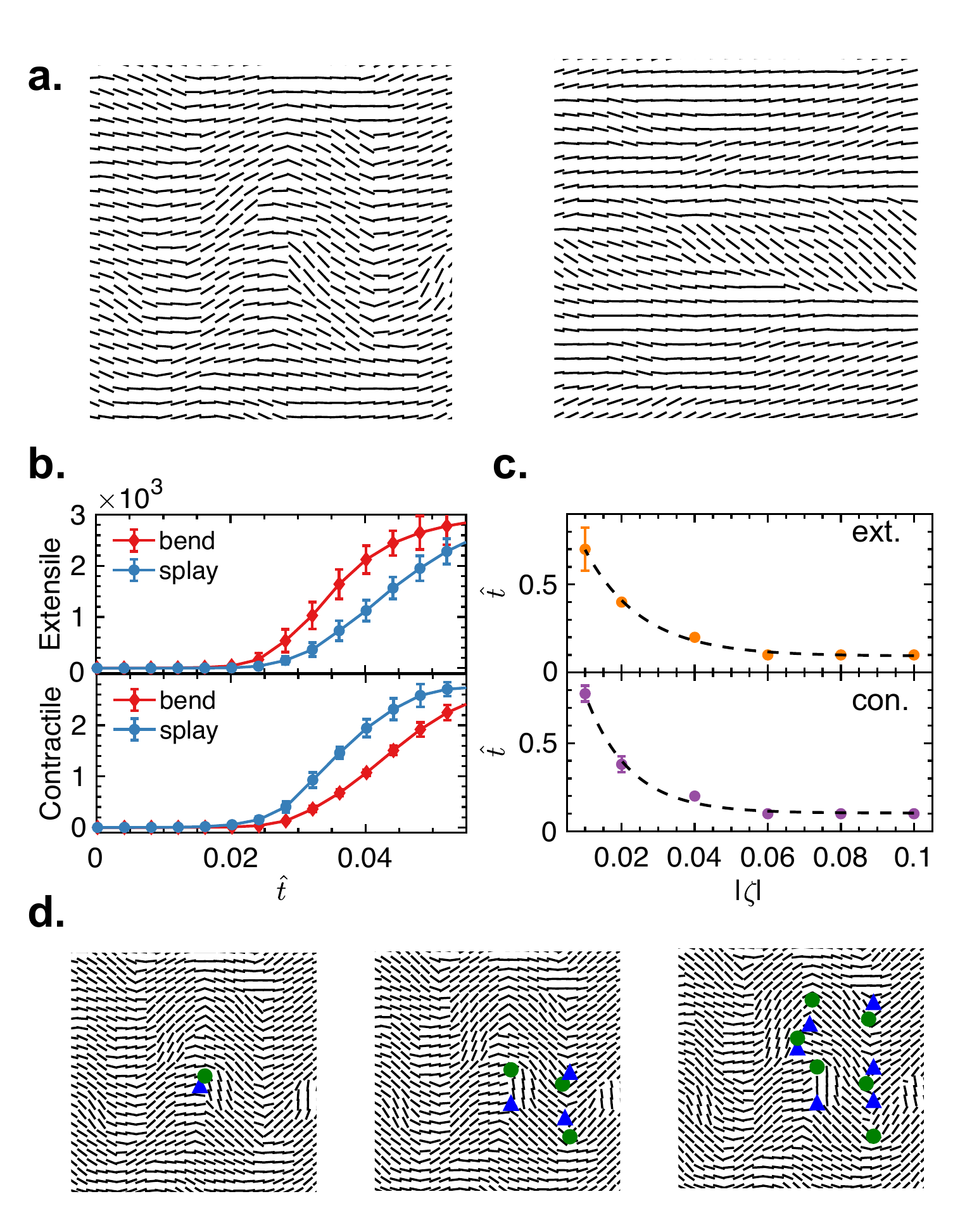}
\caption{\label{fig:bend and splay}%
{\bf Bend-splay instabilities of a cell monolayer.} (a) Tissue nematic field, $\bm{ Q}= \sum_i \bm{Q}_i \phi_i$, showing typical bend (left) and splay (right) instabilities of an initially homogeneous state for extensile ($\zeta>0$) and contractile ($\zeta<0$) activities, respectively. Parameters used here: $\zeta = \pm 0.01$, $\gamma = 0.05$. (b) Evolution of the total bend, $b$, and splay, $s$, given by eqn \eqref{bendsplay} of the tissue nematic field, $\bm{Q}$, as a function of time for extensile (top) and contractile activities (bottom). Averaged over 5 simulations with $\zeta = \pm 0.02$ and $\gamma=0.05$. (c) Number of time steps (rescaled by $1/RK$) before creation of the first defect pair for different extensile (top) and contractile activities (bottom). The dashed line is the exponential fit. Mean$\pm$std from 5 simulations with $\gamma=0.1$. (d) Snapshots of simulation demonstrating unzipping of walls by topological defects during bend instability for an extensile activity. Green circle: $+1/2$-defect, blue triangle: $-1/2$-defect. The simulation was performed with $\zeta=0.01$ and $\gamma=0.05$.}
 \end{figure}

Continuum theory shows that when a nematic system is active, it is prone to bend or splay instability, which depends on the sign of the activity parameter~\cite{simha2002hydrodynamic, ramaswamy2007active, thampi2014instabilities}. Let us define bend and splay measure as follows:
\begin{equation}\label{bendsplay}
b =  \int \dd \vec{x}\, (\vec{n} \wedge \vec{\nabla} \wedge \vec{n})^2/2, \, \, \, s = \int \dd \vec{x}\, (\vec{\nabla} \cdot \vec{n})^2/2.
\end{equation}
Initializing a system from a quiescent state, we observe that after some transient relaxation time, the tissue nematic field, $\bm{Q}$, develops non-zero bend (splay) when the cell-cell interaction is extensile (contractile), as demonstrated in Fig.~\ref{fig:bend and splay}a,b. This initial instability then leads to the formation of walls (Fig.~\ref{fig:bend and splay}a) -- elongated distortions in the director field -- which are also predicted by the continuum theory~\cite{ramaswamy2007active,giomi2011excitable,marchetti2013hydrodynamics}.
 
At longer times, we observe the spontaneous creation of pairs of topological defects that lead to the unzipping of walls in the orientation field for both bend and splay instabilities, see Fig. \ref{fig:bend and splay}d. In order to quantify the timescale associated with the formation of these instabilities, we measure the time before the creation of the first defect pair for different values of the activity, $\zeta$ (Fig.~\ref{fig:bend and splay}c). We observe that as the activity increases, the time to the formation of first defect pair decreases exponentially, which is consistent with the predictions from continuum theory on the exponential growth of the instabilities in time with the time constant proportional to $1/|\zeta|$, where $\zeta$ is the activity coefficient~\cite{giomi2011excitable,thampi2014instabilities}.

\subsection{Analysis of the instabilities at linear order\label{sec:LSA}}
In this section, we show how the dynamics presented in eqns~\eqref{eq:theta} and~\eqref{eq:torques} capture the bend and splay instabilities observed in continuum systems of active nematics by performing a linear stability analysis.
To keep the description analytically tractable, we make several assumptions that allow us to derive a continuum approximation of our multi-agent model in the double limit where both interfaces and cells are small. We then consider the evolution of this continuum description from a quiescent state at linear order and obtain stability conditions for the system.

\begin{figure}[ht!]
    \centering \includegraphics[width=1\linewidth]{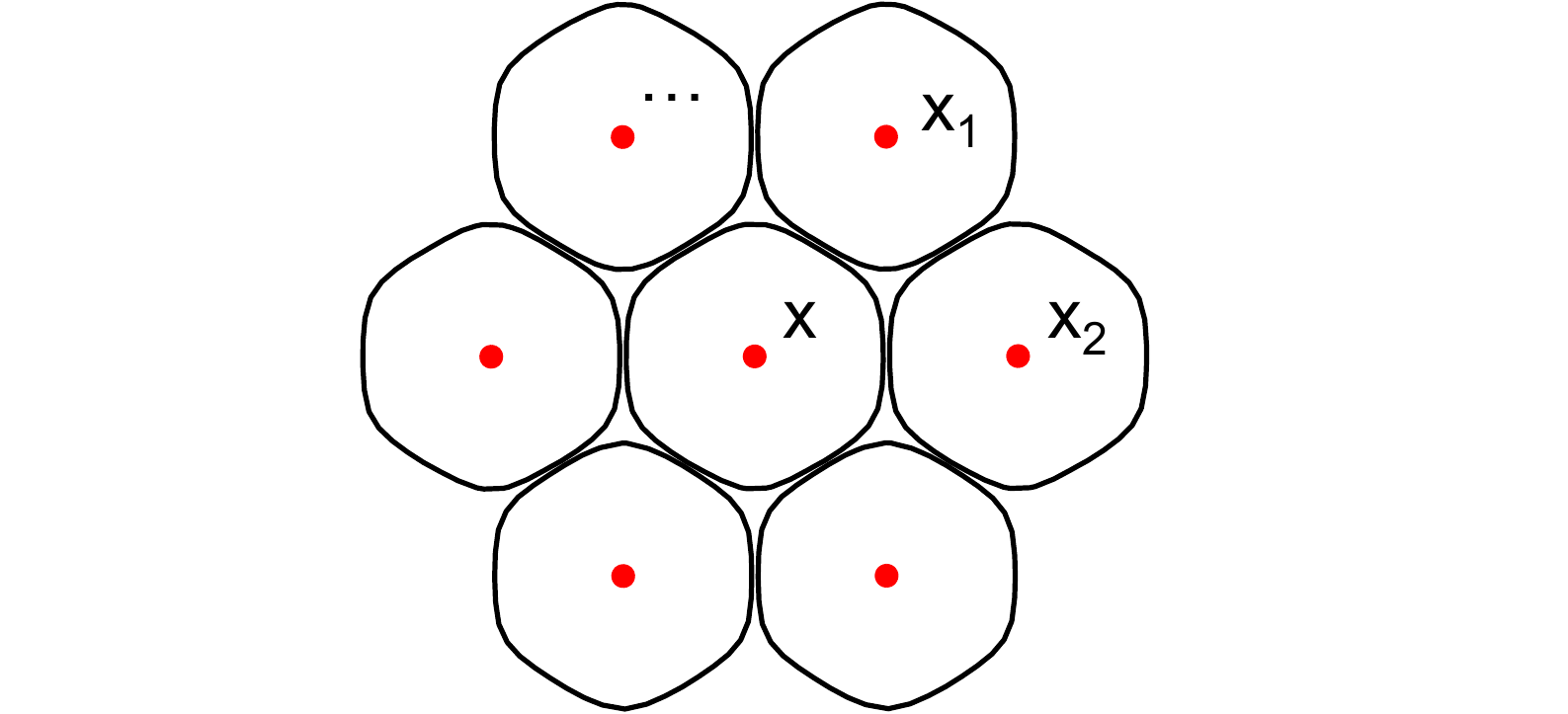}
    \caption{\label{fig:honey}%
        A cell and its neighbours in a honeycomb configuration showing approximate hexagonal shape.
        Additional surrounding cells are not shown.
        The locations of the centre of the first cell $\textbf{\textit{x}}_0$ and of two of its neighbours $\vec{x}_i = \vec{x}_0 + \delta \vec{m}_{i-1}$ are indicated.
    }
\end{figure}

Our procedure is as follows. First, note that in the absence of activity, i.e. $\zeta=0$, the epithelium relaxes to a honeycomb lattice where all cells take an approximate hexagonal form (Fig.~\ref{fig:honey}).
Consider then a cell with phase-field $\phi_0$ centred at $\vec{x}$ with neighbouring cells located at $\vec{x}_i = \vec{x}_0 + \delta \vec{m}_{i-1}$ for $i=1, \ldots, 6$, where $\delta$ is the distance between the centres of two neighbouring cells, and $\vec{m}_i = \left(\cos \left( i \pi / 3\right), \sin \left( i \pi / 3\right) \right)$.
We then assume that the widths of the cells interfaces are much smaller than the cells sizes, i.e. $\lambda \ll \delta$, such that the cells take an hexagonal form with $\delta \approx \sqrt 3 R$.
In this limit, all the integrals over the phase fields can easily be performed by separating the directions perpendicular and parallel to the interfaces.
Assuming constant field along the parallel direction and equilibrium interface profiles $\phi_\pm (\vec{x} - \vec{x}_0) = \frac 1 2 \left(1 \pm \tanh\left( \vec{x} / l \right) \right)$ along the perpendicular direction, we can use the explicit integrals
\[
     \int \dd \vec{x}\, \phi_\pm(\vec{x}) \phi_\mp(\vec{x}) = l/2, \quad
     \int \dd \vec{x}\, \phi_\pm(\vec{x}) \partial_x \phi_\mp(\vec{x}) = \mp 1/2,
\]
to obtain the following expressions
\[
     \int \dd \vec{x}\, \phi_i \phi_0 \approx \frac{\delta l}{2 \sqrt 3}, \qquad \text{and} \quad
     \int \dd \vec{x}\, \phi_i \vec{\nabla} \phi_0 \approx - \frac{\delta \vec{m}_i}{2 \sqrt 3},
\]
where $\delta / \sqrt 3$ is the length of the interface between cells $0$ and $i$. Additionally, we have used the fact that the gradients are perpendicular to the interfaces, i.e. $\phi_i \vec{\nabla} \phi_0 \approx \phi_i \vec{m}_i (\vec{m}_i \cdot \vec{\nabla}) \phi_0$.
Note that even though we assume that the interfaces take their equilibrium profiles, the interface length, $l$, will be in principle different from $\lambda$ because we consider a confluent state where the steric interactions between cells cannot be ignored.
Once such integrals over the individual cells have been computed, we can consider the limit where cells are small, i.e. $\lambda \ll \delta \ll 1$.
We rescale all the physical parameters that multiply terms proportional to integrals over the cells interfaces by the interface length, i.e. we replace $K \to K / \delta$, $J \to J / \delta$, and $\xi \to \xi \delta$, and rewrite $\bm{Q}_i$ and $\vec{v}_i$ as fields.
Performing an expansion to first order in $\delta$ finally allows us to obtain expressions for the tissue level fields in the continuum limit and perform a linear stability analysis.

We first apply this procedure to better understand the effect of the torques in eqn~\eqref{eq:torques}.
Considering the cell $i=0$ located at $\vec{x}$, we can write
\begin{align*}
    \tau_0 &= \frac 1 \lambda \int \dd \vec{x}\, \bm{Q}_0 \phi_0  \wedge \bm{Q} \approx \frac l \lambda \frac {\delta} {2 \sqrt 3} \sum_{i=1}^6 \bm{Q}_0 \wedge \bm{Q}_i \\ 
    &= \frac l \lambda \frac \delta {2 \sqrt 3} \sum_{i=1}^6 \bm{Q}(\vec{x}) \wedge \bm{Q}(\vec{x} + \delta \vec{m}_i) \\
    &= \frac l \lambda \frac {\sqrt 3}{2} \Delta \theta(\vec{x}) \delta^3  + \mathcal O (\delta^5),
\end{align*}
where we have set $Q_{xx} = \cos 2\theta$ and $Q_{xy} = \sin 2\theta$.
This form is similar to the elastic term of continuum theories of liquid crystals in the one-elastic constant approximation, and shows that the torque $\tau_0$ does indeed have the effect of an elastic restoring force.
We can similarly rewrite the vorticity torque as
\begin{align*}
    \omega_0 &= \int \dd \vec{x}\,  \vec{v} \wedge \vec{\nabla} \phi_0 = - \frac \delta {2 \sqrt 3} \sum_{i=1}^6 \vec{v}_i \wedge \vec{m}_i \\ 
    &\approx - \frac \delta {2\sqrt 3} \sum_{i=1}^6 \vec{v}(\vec{x} + \delta \vec{m}_i) \wedge \vec{m}_i \\
    &=  \frac {\sqrt 3} 2 (\partial_x v_y(\vec{x}) - \partial_y v_x(\vec{x}) \delta^2 + \mathcal O (\delta^4).
\end{align*}
This shows that the torque $\omega_i$ indeed measures the local tissue vorticity around a cell.
Finally, we consider the total force on the interface and, neglecting the pressure term, obtain the following expansion
\begin{align*}
\vec{F}_0^{\text{int.}} &= \zeta \int \dd \vec{x} \, \bm{Q} \cdot \vec{\nabla} \phi_0 \approx - \frac {\zeta \delta} {2\sqrt 3} \sum_{i=1}^6 \bm{Q}_i \cdot \vec{m}_i \\
&= - \frac {\zeta \delta} {2\sqrt 3} \sum_{i=1}^6 \bm{Q}(\vec x + \delta \vec{m}_i) \cdot \vec{m}_i \\
&= - \frac {\sqrt 3 \zeta} 2\,\begin{pmatrix} \partial_x Q_{xx} +\partial_y Q_{xy} \\ \partial_x Q_{xy} - \partial_y Q_{xx}\end{pmatrix} \delta^2 + \mathcal O (\delta^4),
\end{align*}
which is similar to the local force arising from the active stress term in continuum theories of liquid crystals.

We can now obtain the linear stability condition for the homogeneous state $\theta=0$ by using the force balance equation $\vec{v} = \vec{F}_0^{\text{int.}}/\xi$ and expand eqn~\eqref{eq:theta} to linear order in $\theta$.
This gives us
\[
\partial_t \theta = \delta^2 \left[\frac l \lambda \frac {\sqrt 3 K}{2} \Delta +  \frac{3 J \zeta}{2 \xi} (\partial_x^2 - \partial_y^2)\right] \theta + \mathcal O(\delta^4, \theta^2),
\]
where we have neglected terms of order $\delta^4$ and $\theta^2$.
For extensile (contractile) systems, it is easy to consider perturbations homogeneous under translation in the $y$ direction ($x$ direction) and obtain the linear stability condition, $-\zeta_c<\zeta<\zeta_c$, where the critical activity is given by
\[
    \zeta_c = \frac l \lambda \frac{\xi K}{3 R J} \propto \frac{\xi K}{J}.
\]
Hence, we see that the system shows bend and splay instabilities when $\zeta > \zeta_c$ and $\zeta < \zeta_c$, respectively.

\begin{figure}[htb!]
    \includegraphics[width=1\linewidth]{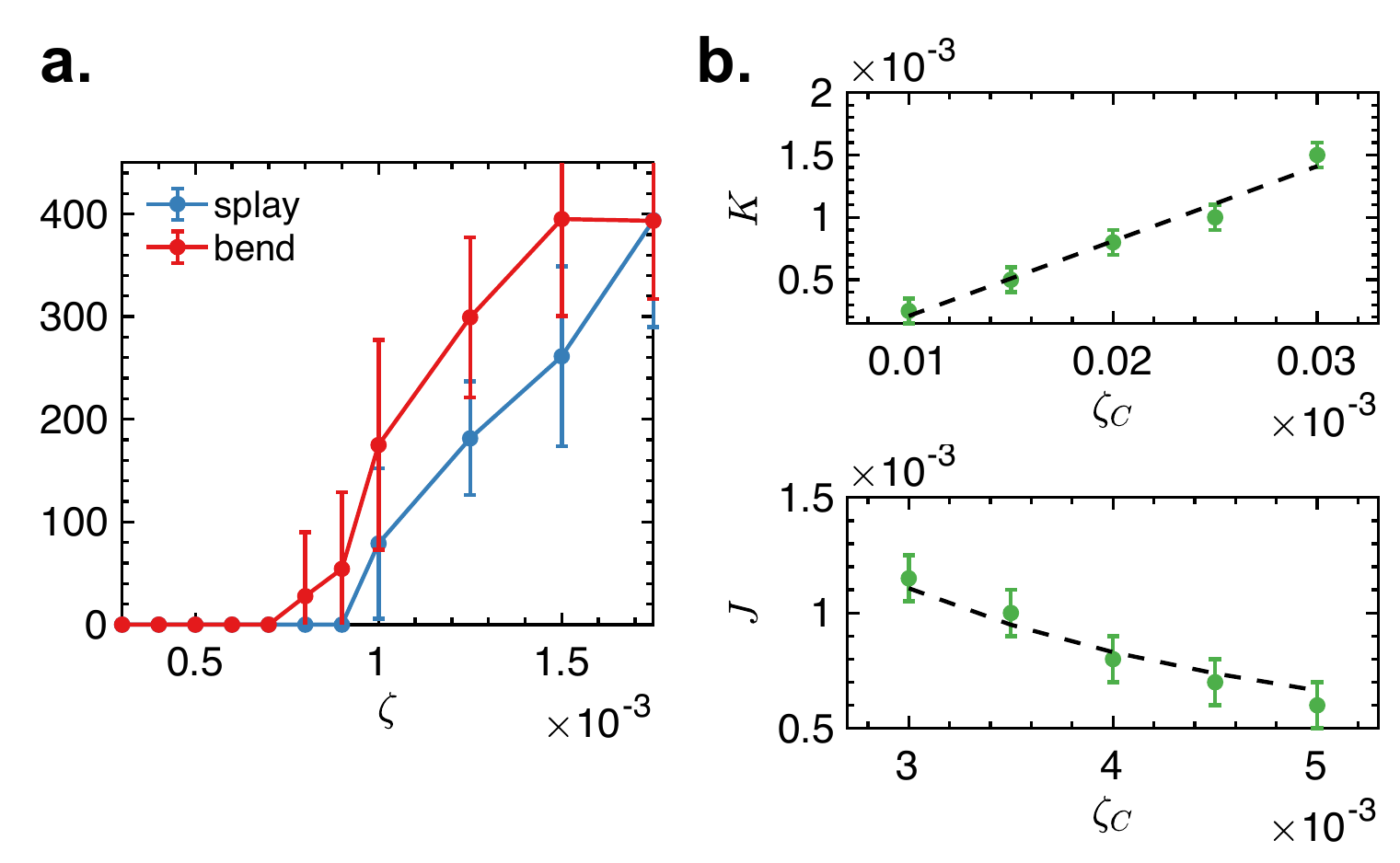} 
    \caption{\label{fig:zetac} {\bf Activity threshold for the instabilities.} (a) Bend and splay (eqn \eqref{bendsplay}) after 10000 steps for different values of activity $\zeta$, with $\gamma = 0.05$, $K = 2\times 10^{-5}$ and $J = 4 \times 10^{-3}$. Mean$\pm$std over 5 simulations. (b) Dependence of the critical activity, $\zeta_C$, on the nematic elastic parameter, $K$, (for $J = 4 \times 10^{-3}$) and nematic flow alignment strength, $J$, (with $K = 2 \times 10^{-5}$) for $\gamma = 0.05$. Dashed lines represent scaling: $\zeta_C \propto K$ and $\zeta_C \propto 1/J$, as predicted from the linear stability analyses.}
\end{figure}

We examine the predictions of the linear stability analysis by performing simulations for different values of the activity parameter, $\zeta$, of the nematic elastic parameter, $K$, and of the nematic flow alignment strength, $J$.
For each of the simulations we measure the average bend and splay defined by eqn.~\eqref{bendsplay} over the last 5000 timesteps.
Fig. \ref{fig:zetac}a shows the average bend and splay as a function of activity for fixed $K$ and $J$, confirming the existence of an activity threshold.
It is apparent that beyond a certain activity value, the system develops non-zero bend and splay as predicted by linear stability analysis, both in the extensile and contractile case.
We then estimate the critical activity $\zeta_C$ from simulations by finding the first value of $\zeta$ for which bend deviates substantially, i.e. $b \geq 100$, from zero at the end of the simulation (last 5000 time steps) and examine how this critical value depends on the elastic constant $K$ and flow-alignment constant $J$. Fig. \ref{fig:zetac}b shows the critical activity $\zeta_C$ as a function of $K$ (top) and $J$ (bottom). The dashed lines demonstrates the expected scaling as predicted from the linear stability analyses, i.e., $\zeta_C \propto K$ and $\zeta_C \propto 1/J$. Together, these results establish the existence of a well-defined critical activity, above which a collection of active deformable cells exhibit bend/splay instabilities.
\begin{figure*}[htb!]
    \includegraphics[width=1\linewidth]{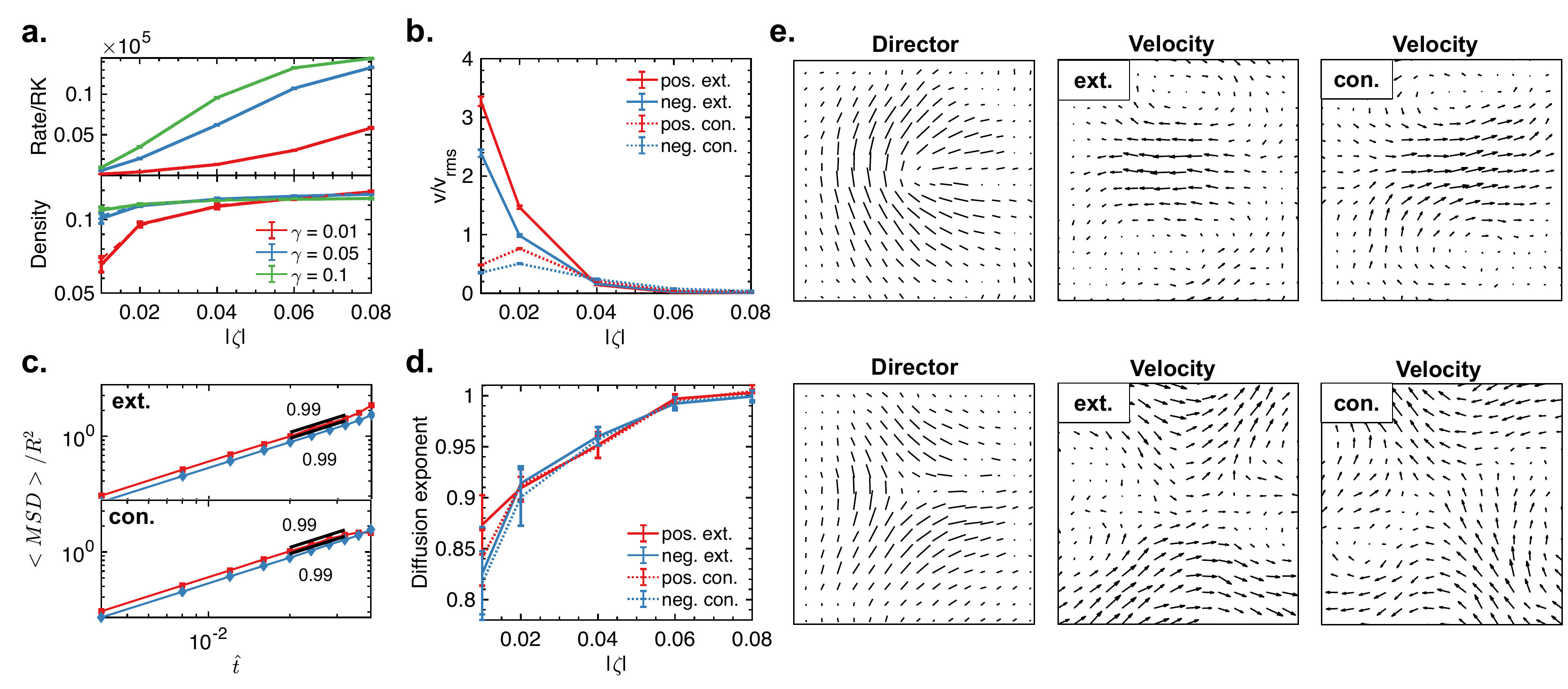} 
    \caption{\label{fig:vrms}%
    {\bf Topological defect statistics.}
    	(a) Rescaled rate of creation of defect pairs (top) and density of defects (bottom) as a function of the activity, $|\zeta|$, for different values of the elastic constant, $\gamma$. Solid line -- extensile, dotted -- contractile (dashed lines not visible since extensile and contractile curves overlap). Mean$\pm$std from 5 simulations.
	(b) Defect speed divided by mean root-mean-square velocity of the entire monolayer for positive (red) and negative (blue) defects for extensile (solid) and contractile (dotted) activities. Here, $\gamma = 0.1$. Mean$\pm$std from 5 simulations.
        (c) Mean-squared-distance (rescaled by squared cell radius, $R^2$) travelled by defect from its creation point as a function of time for positive (red) and negative (blue) defects with $\zeta=\pm 0.01$ (left) and $\zeta=\pm 0.06$ (right), and $\gamma = 0.1$. Top row: extensile, bottom row: contractile system. The time is rescaled by $1/RK$. Mean$\pm$std from all defects in one simulation.
        (d) Diffusion exponent for $\gamma=0.1$ as a function of activity for extensile (solid) and contractile (dotted) defects. Mean$\pm$std from 5 simulations. 
        (e) Average properties of $+1/2$-defects (top) and $-1/2$-defects (bottom) in an extensile and contractile systems with $\zeta=\pm0.04$ and $\gamma = 0.1$. Average over time and over 5 simulations.
    }
\end{figure*}

\subsection{Active turbulence: defects characteristics\label{sec:def}}
As expected from continuum theory of liquid crystals, the hydrodynamic instability of the ordered state and the subsequent defect pair nucleation is followed, at longer times and for high enough activity, by the emergence of active turbulence~\cite{giomi2013defect}. The chaotic flows of active cellular systems are often interleaved with dynamic organisation of topological defects~\cite{saw2017topological,kawaguchi2017topological,blanch2018turbulent}. As such, we next look at the rate of defect pair creation and the density of defects for different values of activity, $\zeta$, and elasticity, $\gamma$, shown in Figure~\ref{fig:vrms}a. In line with the predictions of continuum theories of active nematics~\cite{giomi2013defect, thampi2014instabilities}, both values of rate and density increase for larger activities in both contractile and extensile systems until they saturate. The rate of saturation depends on the elasticity, i.e. larger values of $\gamma$ lead to more rapid increase in defect nucleation rate and a faster saturation of the defect density.

It is well established that although topological defects in active nematics appear and annihilate in pairs with positive and negative half-integer charge, the defects with different charge show distinct dynamic behaviour. In particular, while negative defects are three-fold symmetric and lack self-propulsion, the positive, comet-shaped, defects in active systems are endowed with self-propulsive velocity along their comet-head (-tail) for extensile (contractile) activities.  

To characterize the dynamics of positive and negative half-integer defects we first measured the average speed of defects normalised by root-mean-squared velocity of the flow in the entire domain, which demonstrate higher velocities of positive defects at low activities (Fig. \ref{fig:vrms}b). At low activities, both positive and negative half-integer defects show larger velocities in extensile systems compared to the contractile one. At higher activities, however, the normalised defect velocities, for both positive and negative half-integer defects and in both extensile and contractile systems, drop to the same level. To study this trend more closely, we calculate the mean squared distance travelled from the time of defect creation for each defect as a function of time (Fig. \ref{fig:vrms}c) and obtain the corresponding diffusion exponent at long times (Fig. \ref{fig:vrms}d). When the activity is relatively low, the defect density is lower and positive defects show larger diffusion exponents, indicating the impact of self-propulsion on their dynamics. However, when the activity is increased and the density of the defects approaches its saturation values (Fig. \ref{fig:vrms}a), the diffusion constants of both positive and negative defects become equal and tend to $1$. Similar behaviour is observed for different values of elasticity, $\gamma$, and remains the same in both extensile and contractile systems. This is in contrast to the predictions of continuum theories~\cite{thampi2014instabilities} and observations in bacterial colonies~\cite{meacock2021bacteria} and subcellular filaments-motor protein mixtures~\cite{Sanchez2012,kumar2018tunable}, where positive and negative half-integer defects show distinct propulsive and diffusive behaviours, respectively. Remarkably, however, the diffusive behaviour for both defects, observed in our simulations at high activities, is in agreement with the experimental observations on Human Bronchial Cells (HBC)~\cite{blanch2018turbulent} and recent phase-field modelling with distinct mechanisms of active driving~\cite{wenzel2021multiphase}, suggesting that such diffusive defect dynamics are generic features of deformable active particles at high activities, where the nematic order is an emergent property and defects motion is dominated by the interaction between defects at long times.
\begin{figure*}[htb!]
    \includegraphics[width=1\linewidth]{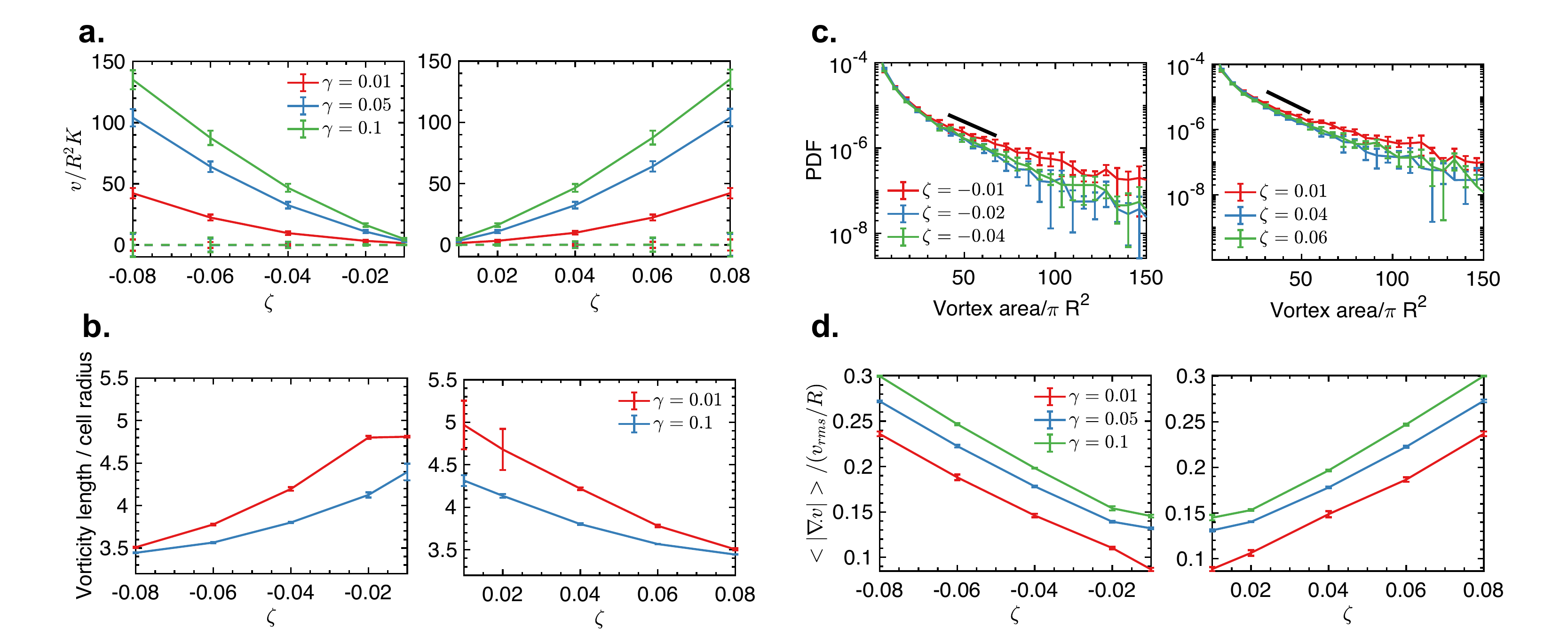} 
    \caption{\label{fig:corrs}
    {\bf Global flow features of the active turbulence in the monolayer.}
    (a) Rescaled root-mean-square velocity, $\sqrt{v_x^2+v_y^2}$, (solid) and sum of both components, $v_x+v_y$, (dashed) plotted for different values of activity, $\zeta$, and elasticity, $\gamma$. Averaged over the whole domain and time (5 realisations). (b) Dependence of the vorticity length defined as the location of the minimum of the vorticity autocorrelation function as a function of $\zeta$ for different values of $\gamma$. Mean$\pm$std from 5 simulations. (c) Distribution of vortex areas, defined by Okubo-Weiss parameter, for different values of activities, $\zeta$, with $\gamma=0.1$. Averaged over 5 realisations. The $x$-axis  is rescaled by $\pi R^2$ and the black solid-line marks a possible exponential scaling at intermediate vortex sizes. (d) Absolute value of the divergence as a function of activity, $\zeta$, rescaled by $v_\text{rms}/R$. Averaged over 5 simulations.}
\end{figure*}

To further complement the analyses of the defect dynamics and highlight the structural differences between the different defect types, we also calculated the average flows around positive and negative defects since the flow patterns around defects are crucial in determining the dynamics of active turbulence~\cite{giomi2013defect} and have been recently shown to play a role in the mechanical properties of epithelial monolayers~\cite{saw2017topological,kawaguchi2017topological,doostmohammadi2021physics}.
We extract the average properties of defects in the case of extensile interaction $\zeta>0$ and obtain the local properties for positive defects (top row) and for negative defects (bottom row), as shown in Fig.~\ref{fig:vrms}e for both contractile and extensile defects.
They reproduce faithfully the theoretical predictions from continuum models as well as the experimental observations made in epithelial monolayers~\cite{saw2017topological}.

\subsection{Active turbulence: flow characteristics\label{sec:flow}}
Next, we focus on the global features of the flow field in the monolayer as the active turbulent state is established.

First, we measure both the root-mean-square velocity, $\sqrt{v_x^2+v_y^2}$, and the ensemble average of the sum of both velocity components, $v_x+v_y$. In a confluent epithelium, the total force is approximately zero and the system does not develop any total velocity under periodic boundary conditions. This is demonstrated in Fig. \ref{fig:corrs}a (dashed line) for both extensile and contractile systems. In contrast, the rms-velocity has a finite value and is enhanced with increasing activity for both extensile and contractile systems. More importantly, the generated flow is interleaved with vortical structures. This is quantified by calculating the vorticity-vorticity correlation function that clearly shows the existence of a characteristic vorticity length scale for different activities. Consistent with continuum theories~\cite{thampi2014instabilities}, the vorticity length scale decreases with increasing activity (Fig.~\ref{fig:corrs}b).

In addition to having an intrinsic vorticity length scale that depends on the activity of the system, active turbulence is shown to manifest exponential distribution of vortex sizes, that was first predicted using active nematohydrodynamic equations~\cite{giomi2015geometry}, and was confirmed in epithelial monolayers of Human Bronchial Cells~\cite{blanch2018turbulent}, and in dense suspensions of microtubule-kinesin motor mixtures~\cite{martinez2019selection}. We quantify the vortex areas using Okubo-Weiss criterion~\cite{blanch2018turbulent}, $\mathcal{W}$:
\begin{equation}
\mathcal{W} = -\textup{det} \left[ \vec{\nabla} \vec{v} \right].
\end{equation}
The region is considered to be a vortex if $\mathcal{W} < 0$. Fig. \ref{fig:corrs}c demonstrates the semi-log representation of the total vertex area probability distribution for different values of activity with $\gamma=0.1$. While we see a possible evidence of activity-dependent, exponential scaling at intermediate vortex areas, we could not find a conclusive evidence of the exponential distribution of vortex areas in our simulations.

Additionally, continuum models of epithelial layers often assume the velocity field of the cells to be incompressible~\cite{saw2017topological,duclos2018spontaneous,perez2019active,alert2019active}. We directly check this constraint in our cell-based model by calculating the average of the absolute values of the flow divergence, $\vec{\nabla}\cdot \vec{v}$, for different values of $\zeta$ and $\gamma$, as shown in Fig. \ref{fig:corrs}d. Interestingly, as the activity increases, the divergence becomes non-zero, suggesting that the monolayer of cells deviates from an incompressible behavior. Higher values of divergence are observed for larger values of elasticity, $\gamma$. The non-zero divergence further indicates the emergence of dilating-shrinking domains in the system, which we explore next.

\subsection{Spontaneous gap formation\label{sec:gap}}
\begin{figure}[h]
    \includegraphics[width=\linewidth]{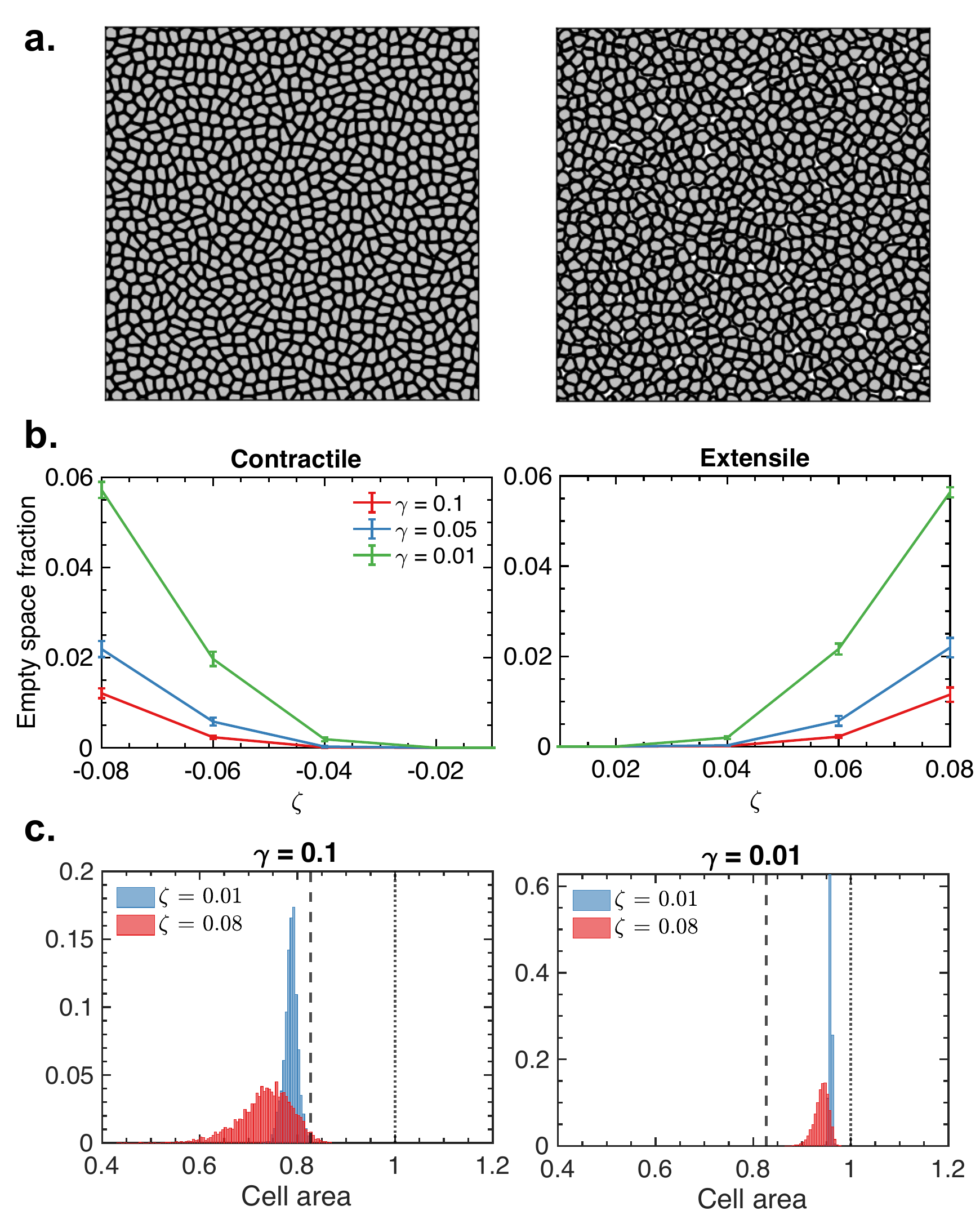} 
    \caption{\label{fig:gaps}%
        Higher activity values lead to heterogeneity in cell densities, eventually leading to the formation of highly dense regions and empty spaces, i.e. gaps. (a) Plots of $\phi_{tot}(t) = \sum_{i} \phi_i(t)$ for $\zeta = 0.01$ with $\gamma = 0.1$ (left) and $\zeta = 0.08$ with $\gamma = 0.01$ (right) at the end of the simulation. Black lines represent cell contours, grey -- interior of the cell, and white -- exterior. (b) Fraction of empty space on the grid as a function of time for contractile (left) and extensile (right) activities. The empty space was evaluated at the end of simulation (15000 time steps) and averaged over 5 simulations. (c) Cell area distribution (normalized) for different values of activity, $\zeta$, and elasticity, $\gamma$. Dashed vertical line corresponds to the area of hexagon, $3\sqrt{3}R^2/2$, and the dotted line to the area of circle, $\pi R^2$. The $x$-axis is rescaled by $\pi R^2$. 
    }
\end{figure}

While physical properties of confluent monolayers have been intensely investigated, recent modelling and {\it in vivo} experiments have begun to unravel the importance of extracellular spaces on structural properties of the cell layers~\cite{kim2021embryonic}. Interestingly, we find that the emergence of extracellular spaces can be governed by the activity level of the cells: at a given cell density, for sufficiently high activities the confluency condition is broken and extracellular spaces spontaneously emerge (Fig. \ref{fig:gaps}a).

To quantitatively characterize the spontaneous gap formation, we measure the fraction of empty space within the cell layer. To this end, in the phase-field model, we use the threshold method. We calculate the total cell density by summing over all cells:
\begin{equation}
\phi_{tot} = \sum_i^N \phi_i.
\end{equation}
We then assume that the grid points with $\phi_{tot}<\phi_0$ correspond to empty space. The value of $\phi_0=0.2$ is chosen by qualitatively identifying the gaps. 

Interestingly, measurement of the empty space fraction over simulation time shows that after their formation the fraction reaches a constant value throughout the simulations. Fig. \ref{fig:gaps}b demonstrates the steady-state fraction of empty space for different values of activity and elasticity, showing an elasticity-dependent threshold for activity, above which extracellular spaces are formed. Beyond this point, the empty space fraction increases as the activity is increased. Similar gap formation behavior is observed for both extensile and contractile systems (Fig.~\ref{fig:gaps}b), indicating that the spontaneous emergence of extracellular spaces is a generic feature of an active multicellular layer. 

Furthermore, the crossover from the confluent to non-confluent state is accompanied by a significant alteration of the individual cell features. This is best quantified by measuring the distribution of cell areas within the confluent and non-confluent states of the monolayer (Fig.~\ref{fig:gaps}c). When the monolayer is confluent (left panel in Fig.~\ref{fig:gaps}c), the cells obtain more hexagonal shape, i.e. on average closer to the dashed line, indicating the area of a hexagon. However, when the confluency is broken (right panel in Fig.~\ref{fig:gaps}c), the cells have more circular shape, i.e. on average closer to the dotted line, indicating the area of a circle, and the cell distributions are narrower compared to the confluent case. Furthermore, as the activity increases, the distribution broadens significantly, indicating the emergent activity-induced heterogeneity in cell areas, and is true for both confluent case and monolayers with gaps.
\section{Conclusions\label{sec:conc}}
We have presented a phase-field model of cellular monolayers that bridges cell-level interaction forces to the active nematohydrodynamic behavior at the multicellular level. We do so by introducing a minimal form of local dipolar interaction between cells and show that this simple formulation reproduces phenomenology of the continuum active nematohydrodynamic approach. In particular, by performing numerical simulations, we show the emergence of bend and splay instabilities for extensile and contractile interactions. At later times, the system develops active turbulence and the formation of topological defects. We quantify the observed features of flows and defects for both extensile and contractile systems, and show that they are in agreement with the experimental observations.

Furthermore, we find that when activity and elasticity of individual cells is sufficiently high, the monolayer flow field deviates from its incompressible behavior. Concomitantly, we demonstrate that this deviation of the cell layer flow field from incompressibility leads to spontaneous formation of gaps within the confluent layers. We observe larger fraction of empty space for higher activity and lower elasticity. Furthermore, we quantify the emergent heterogeneity in the cell areas, showing that cells in confluent monolayer have more hexagonal shapes and higher degree of heterogeneity compared to cells in monolayer with gaps. The spontaneous gap formation is accompanied by more homogeneous circular cell shapes.

Inter-cellular gap formation has been observed experimentally in epithelial cell monolayers, where gaps can reach the length of several cell bodies \cite{zheng2017epithelial}. Appearance of cell-free space has been linked to viscoelastic properties of the substrate \cite{xi2019material}. More recently, Sonam et al. \cite{sonam2022mechanical} have suggested that the spontaneous formation of holes within cell monolayer can occur due to the heterogeneity in substrate stiffness and using a vertex model they show that the gaps can form in the areas close to negative half-integer topological defects, which result in higher tensile stresses within the cell layer. Thus, extending the model to account for substrate heterogeneity will provide a theoretical platform to study the gap formation and its implications on collective cell movement.

\section*{Conflicts of interest}
There are no conflicts to declare.

\section*{Acknowledgements}
A. D. acknowledges funding from the Novo Nordisk Foundation (grant No. NNF18SA0035142), Villum Fonden (Grant no. 29476). A. A. and A. D. acknowledge support from the European Union’s Horizon 2020 research and innovation program under the Marie Sklodowska-Curie grant agreement No. 847523 (INTERACTIONS).

\addcontentsline{toc}{chapter}{Bibliography}
\bibliography{rsc}

\begin{thebibliography}{57}%
\makeatletter
\providecommand \@ifxundefined [1]{%
 \@ifx{#1\undefined}
}%
\providecommand \@ifnum [1]{%
 \ifnum #1\expandafter \@firstoftwo
 \else \expandafter \@secondoftwo
 \fi
}%
\providecommand \@ifx [1]{%
 \ifx #1\expandafter \@firstoftwo
 \else \expandafter \@secondoftwo
 \fi
}%
\providecommand \natexlab [1]{#1}%
\providecommand \enquote  [1]{``#1''}%
\providecommand \bibnamefont  [1]{#1}%
\providecommand \bibfnamefont [1]{#1}%
\providecommand \citenamefont [1]{#1}%
\providecommand \href@noop [0]{\@secondoftwo}%
\providecommand \href [0]{\begingroup \@sanitize@url \@href}%
\providecommand \@href[1]{\@@startlink{#1}\@@href}%
\providecommand \@@href[1]{\endgroup#1\@@endlink}%
\providecommand \@sanitize@url [0]{\catcode `\\12\catcode `\$12\catcode
  `\&12\catcode `\#12\catcode `\^12\catcode `\_12\catcode `\%12\relax}%
\providecommand \@@startlink[1]{}%
\providecommand \@@endlink[0]{}%
\providecommand \url  [0]{\begingroup\@sanitize@url \@url }%
\providecommand \@url [1]{\endgroup\@href {#1}{\urlprefix }}%
\providecommand \urlprefix  [0]{URL }%
\providecommand \Eprint [0]{\href }%
\providecommand \doibase [0]{http://dx.doi.org/}%
\providecommand \selectlanguage [0]{\@gobble}%
\providecommand \bibinfo  [0]{\@secondoftwo}%
\providecommand \bibfield  [0]{\@secondoftwo}%
\providecommand \translation [1]{[#1]}%
\providecommand \BibitemOpen [0]{}%
\providecommand \bibitemStop [0]{}%
\providecommand \bibitemNoStop [0]{.\EOS\space}%
\providecommand \EOS [0]{\spacefactor3000\relax}%
\providecommand \BibitemShut  [1]{\csname bibitem#1\endcsname}%
\let\auto@bib@innerbib\@empty
\bibitem [{\citenamefont {Heisenberg}\ and\ \citenamefont
  {Bella{\"\i}che}(2013)}]{heisenberg2013forces}%
  \BibitemOpen
  \bibfield  {author} {\bibinfo {author} {\bibfnamefont {C.-P.}\ \bibnamefont
  {Heisenberg}}\ and\ \bibinfo {author} {\bibfnamefont {Y.}~\bibnamefont
  {Bella{\"\i}che}},\ }\href@noop {} {\bibfield  {journal} {\bibinfo  {journal}
  {Cell}\ }\textbf {\bibinfo {volume} {153}},\ \bibinfo {pages} {948} (\bibinfo
  {year} {2013})}\BibitemShut {NoStop}%
\bibitem [{\citenamefont {Brugu{\'e}s}\ \emph {et~al.}(2014)\citenamefont
  {Brugu{\'e}s}, \citenamefont {Anon}, \citenamefont {Conte}, \citenamefont
  {Veldhuis}, \citenamefont {Gupta}, \citenamefont {Colombelli}, \citenamefont
  {Mu{\~n}oz}, \citenamefont {Brodland}, \citenamefont {Ladoux},\ and\
  \citenamefont {Trepat}}]{brugues2014forces}%
  \BibitemOpen
  \bibfield  {author} {\bibinfo {author} {\bibfnamefont {A.}~\bibnamefont
  {Brugu{\'e}s}}, \bibinfo {author} {\bibfnamefont {E.}~\bibnamefont {Anon}},
  \bibinfo {author} {\bibfnamefont {V.}~\bibnamefont {Conte}}, \bibinfo
  {author} {\bibfnamefont {J.~H.}\ \bibnamefont {Veldhuis}}, \bibinfo {author}
  {\bibfnamefont {M.}~\bibnamefont {Gupta}}, \bibinfo {author} {\bibfnamefont
  {J.}~\bibnamefont {Colombelli}}, \bibinfo {author} {\bibfnamefont {J.~J.}\
  \bibnamefont {Mu{\~n}oz}}, \bibinfo {author} {\bibfnamefont {G.~W.}\
  \bibnamefont {Brodland}}, \bibinfo {author} {\bibfnamefont {B.}~\bibnamefont
  {Ladoux}}, \ and\ \bibinfo {author} {\bibfnamefont {X.}~\bibnamefont
  {Trepat}},\ }\href@noop {} {\bibfield  {journal} {\bibinfo  {journal} {Nature
  Physics}\ }\textbf {\bibinfo {volume} {10}},\ \bibinfo {pages} {683}
  (\bibinfo {year} {2014})}\BibitemShut {NoStop}%
\bibitem [{\citenamefont {Friedl}\ \emph {et~al.}(2012)\citenamefont {Friedl},
  \citenamefont {Locker}, \citenamefont {Sahai},\ and\ \citenamefont
  {Segall}}]{friedl2012classifying}%
  \BibitemOpen
  \bibfield  {author} {\bibinfo {author} {\bibfnamefont {P.}~\bibnamefont
  {Friedl}}, \bibinfo {author} {\bibfnamefont {J.}~\bibnamefont {Locker}},
  \bibinfo {author} {\bibfnamefont {E.}~\bibnamefont {Sahai}}, \ and\ \bibinfo
  {author} {\bibfnamefont {J.~E.}\ \bibnamefont {Segall}},\ }\href@noop {}
  {\bibfield  {journal} {\bibinfo  {journal} {Nature Cell Biology}\ }\textbf
  {\bibinfo {volume} {14}},\ \bibinfo {pages} {777} (\bibinfo {year}
  {2012})}\BibitemShut {NoStop}%
\bibitem [{\citenamefont {Ladoux}\ and\ \citenamefont
  {M{\`e}ge}(2017)}]{ladoux2017mechanobiology}%
  \BibitemOpen
  \bibfield  {author} {\bibinfo {author} {\bibfnamefont {B.}~\bibnamefont
  {Ladoux}}\ and\ \bibinfo {author} {\bibfnamefont {R.-M.}\ \bibnamefont
  {M{\`e}ge}},\ }\href@noop {} {\bibfield  {journal} {\bibinfo  {journal}
  {Nature Reviews Molecular Cell Biology}\ }\textbf {\bibinfo {volume} {18}},\
  \bibinfo {pages} {743} (\bibinfo {year} {2017})}\BibitemShut {NoStop}%
\bibitem [{\citenamefont {Doostmohammadi}\ and\ \citenamefont
  {Ladoux}(2021)}]{doostmohammadi2021physics}%
  \BibitemOpen
  \bibfield  {author} {\bibinfo {author} {\bibfnamefont {A.}~\bibnamefont
  {Doostmohammadi}}\ and\ \bibinfo {author} {\bibfnamefont {B.}~\bibnamefont
  {Ladoux}},\ }\href@noop {} {\bibfield  {journal} {\bibinfo  {journal} {Trends
  in Cell Biology}\ } (\bibinfo {year} {2021})}\BibitemShut {NoStop}%
\bibitem [{\citenamefont {Duclos}\ \emph {et~al.}(2018)\citenamefont {Duclos},
  \citenamefont {Blanch-Mercader}, \citenamefont {Yashunsky}, \citenamefont
  {Salbreux}, \citenamefont {Joanny}, \citenamefont {Prost},\ and\
  \citenamefont {Silberzan}}]{duclos2018spontaneous}%
  \BibitemOpen
  \bibfield  {author} {\bibinfo {author} {\bibfnamefont {G.}~\bibnamefont
  {Duclos}}, \bibinfo {author} {\bibfnamefont {C.}~\bibnamefont
  {Blanch-Mercader}}, \bibinfo {author} {\bibfnamefont {V.}~\bibnamefont
  {Yashunsky}}, \bibinfo {author} {\bibfnamefont {G.}~\bibnamefont {Salbreux}},
  \bibinfo {author} {\bibfnamefont {J.-F.}\ \bibnamefont {Joanny}}, \bibinfo
  {author} {\bibfnamefont {J.}~\bibnamefont {Prost}}, \ and\ \bibinfo {author}
  {\bibfnamefont {P.}~\bibnamefont {Silberzan}},\ }\href@noop {} {\bibfield
  {journal} {\bibinfo  {journal} {Nature Physics}\ }\textbf {\bibinfo {volume}
  {14}},\ \bibinfo {pages} {728} (\bibinfo {year} {2018})}\BibitemShut
  {NoStop}%
\bibitem [{\citenamefont {Kawaguchi}\ \emph {et~al.}(2017)\citenamefont
  {Kawaguchi}, \citenamefont {Kageyama},\ and\ \citenamefont
  {Sano}}]{kawaguchi2017topological}%
  \BibitemOpen
  \bibfield  {author} {\bibinfo {author} {\bibfnamefont {K.}~\bibnamefont
  {Kawaguchi}}, \bibinfo {author} {\bibfnamefont {R.}~\bibnamefont {Kageyama}},
  \ and\ \bibinfo {author} {\bibfnamefont {M.}~\bibnamefont {Sano}},\
  }\href@noop {} {\bibfield  {journal} {\bibinfo  {journal} {Nature}\ }\textbf
  {\bibinfo {volume} {545}},\ \bibinfo {pages} {327} (\bibinfo {year}
  {2017})}\BibitemShut {NoStop}%
\bibitem [{\citenamefont {Prost}\ \emph {et~al.}(2015)\citenamefont {Prost},
  \citenamefont {J{\"u}licher},\ and\ \citenamefont
  {Joanny}}]{prost2015active}%
  \BibitemOpen
  \bibfield  {author} {\bibinfo {author} {\bibfnamefont {J.}~\bibnamefont
  {Prost}}, \bibinfo {author} {\bibfnamefont {F.}~\bibnamefont {J{\"u}licher}},
  \ and\ \bibinfo {author} {\bibfnamefont {J.-F.}\ \bibnamefont {Joanny}},\
  }\href@noop {} {\bibfield  {journal} {\bibinfo  {journal} {Nature Physics}\
  }\textbf {\bibinfo {volume} {11}},\ \bibinfo {pages} {111} (\bibinfo {year}
  {2015})}\BibitemShut {NoStop}%
\bibitem [{\citenamefont {Doostmohammadi}\ \emph {et~al.}(2018)\citenamefont
  {Doostmohammadi}, \citenamefont {Ign{\'e}s-Mullol}, \citenamefont {Yeomans},\
  and\ \citenamefont {Sagu{\'e}s}}]{doostmohammadi2018active}%
  \BibitemOpen
  \bibfield  {author} {\bibinfo {author} {\bibfnamefont {A.}~\bibnamefont
  {Doostmohammadi}}, \bibinfo {author} {\bibfnamefont {J.}~\bibnamefont
  {Ign{\'e}s-Mullol}}, \bibinfo {author} {\bibfnamefont {J.~M.}\ \bibnamefont
  {Yeomans}}, \ and\ \bibinfo {author} {\bibfnamefont {F.}~\bibnamefont
  {Sagu{\'e}s}},\ }\href@noop {} {\bibfield  {journal} {\bibinfo  {journal}
  {Nature Communications}\ }\textbf {\bibinfo {volume} {9}},\ \bibinfo {pages}
  {1} (\bibinfo {year} {2018})}\BibitemShut {NoStop}%
\bibitem [{\citenamefont {Maroudas-Sacks}\ \emph {et~al.}(2021)\citenamefont
  {Maroudas-Sacks}, \citenamefont {Garion}, \citenamefont {Shani-Zerbib},
  \citenamefont {Livshits}, \citenamefont {Braun},\ and\ \citenamefont
  {Keren}}]{maroudas2020topological}%
  \BibitemOpen
  \bibfield  {author} {\bibinfo {author} {\bibfnamefont {Y.}~\bibnamefont
  {Maroudas-Sacks}}, \bibinfo {author} {\bibfnamefont {L.}~\bibnamefont
  {Garion}}, \bibinfo {author} {\bibfnamefont {L.}~\bibnamefont
  {Shani-Zerbib}}, \bibinfo {author} {\bibfnamefont {A.}~\bibnamefont
  {Livshits}}, \bibinfo {author} {\bibfnamefont {E.}~\bibnamefont {Braun}}, \
  and\ \bibinfo {author} {\bibfnamefont {K.}~\bibnamefont {Keren}},\
  }\href@noop {} {\bibfield  {journal} {\bibinfo  {journal} {Nature Physics}\
  }\textbf {\bibinfo {volume} {17}},\ \bibinfo {pages} {251} (\bibinfo {year}
  {2021})}\BibitemShut {NoStop}%
\bibitem [{\citenamefont {Doostmohammadi}\ \emph {et~al.}(2015)\citenamefont
  {Doostmohammadi}, \citenamefont {Thampi}, \citenamefont {Saw}, \citenamefont
  {Lim}, \citenamefont {Ladoux},\ and\ \citenamefont
  {Yeomans}}]{doostmohammadi2015celebrating}%
  \BibitemOpen
  \bibfield  {author} {\bibinfo {author} {\bibfnamefont {A.}~\bibnamefont
  {Doostmohammadi}}, \bibinfo {author} {\bibfnamefont {S.~P.}\ \bibnamefont
  {Thampi}}, \bibinfo {author} {\bibfnamefont {T.~B.}\ \bibnamefont {Saw}},
  \bibinfo {author} {\bibfnamefont {C.~T.}\ \bibnamefont {Lim}}, \bibinfo
  {author} {\bibfnamefont {B.}~\bibnamefont {Ladoux}}, \ and\ \bibinfo {author}
  {\bibfnamefont {J.~M.}\ \bibnamefont {Yeomans}},\ }\href@noop {} {\bibfield
  {journal} {\bibinfo  {journal} {Soft Matter}\ }\textbf {\bibinfo {volume}
  {11}},\ \bibinfo {pages} {7328} (\bibinfo {year} {2015})}\BibitemShut
  {NoStop}%
\bibitem [{\citenamefont {Rossen}\ \emph {et~al.}(2014)\citenamefont {Rossen},
  \citenamefont {Tarp}, \citenamefont {Mathiesen}, \citenamefont {Jensen},\
  and\ \citenamefont {Oddershede}}]{rossen2014long}%
  \BibitemOpen
  \bibfield  {author} {\bibinfo {author} {\bibfnamefont {N.~S.}\ \bibnamefont
  {Rossen}}, \bibinfo {author} {\bibfnamefont {J.~M.}\ \bibnamefont {Tarp}},
  \bibinfo {author} {\bibfnamefont {J.}~\bibnamefont {Mathiesen}}, \bibinfo
  {author} {\bibfnamefont {M.~H.}\ \bibnamefont {Jensen}}, \ and\ \bibinfo
  {author} {\bibfnamefont {L.~B.}\ \bibnamefont {Oddershede}},\ }\href@noop {}
  {\bibfield  {journal} {\bibinfo  {journal} {Nature Communications}\ }\textbf
  {\bibinfo {volume} {5}},\ \bibinfo {pages} {1} (\bibinfo {year}
  {2014})}\BibitemShut {NoStop}%
\bibitem [{\citenamefont {Doostmohammadi}\ \emph {et~al.}(2016)\citenamefont
  {Doostmohammadi}, \citenamefont {Adamer}, \citenamefont {Thampi},\ and\
  \citenamefont {Yeomans}}]{doostmohammadi2016stabilization}%
  \BibitemOpen
  \bibfield  {author} {\bibinfo {author} {\bibfnamefont {A.}~\bibnamefont
  {Doostmohammadi}}, \bibinfo {author} {\bibfnamefont {M.~F.}\ \bibnamefont
  {Adamer}}, \bibinfo {author} {\bibfnamefont {S.~P.}\ \bibnamefont {Thampi}},
  \ and\ \bibinfo {author} {\bibfnamefont {J.~M.}\ \bibnamefont {Yeomans}},\
  }\href@noop {} {\bibfield  {journal} {\bibinfo  {journal} {Nature
  Communications}\ }\textbf {\bibinfo {volume} {7}},\ \bibinfo {pages} {1}
  (\bibinfo {year} {2016})}\BibitemShut {NoStop}%
\bibitem [{\citenamefont {Blanch-Mercader}\ \emph {et~al.}(2018)\citenamefont
  {Blanch-Mercader}, \citenamefont {Yashunsky}, \citenamefont {Garcia},
  \citenamefont {Duclos}, \citenamefont {Giomi},\ and\ \citenamefont
  {Silberzan}}]{blanch2018turbulent}%
  \BibitemOpen
  \bibfield  {author} {\bibinfo {author} {\bibfnamefont {C.}~\bibnamefont
  {Blanch-Mercader}}, \bibinfo {author} {\bibfnamefont {V.}~\bibnamefont
  {Yashunsky}}, \bibinfo {author} {\bibfnamefont {S.}~\bibnamefont {Garcia}},
  \bibinfo {author} {\bibfnamefont {G.}~\bibnamefont {Duclos}}, \bibinfo
  {author} {\bibfnamefont {L.}~\bibnamefont {Giomi}}, \ and\ \bibinfo {author}
  {\bibfnamefont {P.}~\bibnamefont {Silberzan}},\ }\href@noop {} {\bibfield
  {journal} {\bibinfo  {journal} {Physical Review Letters}\ }\textbf {\bibinfo
  {volume} {120}},\ \bibinfo {pages} {208101} (\bibinfo {year}
  {2018})}\BibitemShut {NoStop}%
\bibitem [{\citenamefont {Lin}\ \emph {et~al.}(2021)\citenamefont {Lin},
  \citenamefont {Zhang}, \citenamefont {Bi}, \citenamefont {Li},\ and\
  \citenamefont {Feng}}]{lin2021energetics}%
  \BibitemOpen
  \bibfield  {author} {\bibinfo {author} {\bibfnamefont {S.-Z.}\ \bibnamefont
  {Lin}}, \bibinfo {author} {\bibfnamefont {W.-Y.}\ \bibnamefont {Zhang}},
  \bibinfo {author} {\bibfnamefont {D.}~\bibnamefont {Bi}}, \bibinfo {author}
  {\bibfnamefont {B.}~\bibnamefont {Li}}, \ and\ \bibinfo {author}
  {\bibfnamefont {X.-Q.}\ \bibnamefont {Feng}},\ }\href@noop {} {\bibfield
  {journal} {\bibinfo  {journal} {Communications Physics}\ }\textbf {\bibinfo
  {volume} {4}},\ \bibinfo {pages} {1} (\bibinfo {year} {2021})}\BibitemShut
  {NoStop}%
\bibitem [{\citenamefont {Saw}\ \emph {et~al.}(2017)\citenamefont {Saw},
  \citenamefont {Doostmohammadi}, \citenamefont {Nier}, \citenamefont
  {Kocgozlu}, \citenamefont {Thampi}, \citenamefont {Toyama}, \citenamefont
  {Marcq}, \citenamefont {Lim}, \citenamefont {Yeomans},\ and\ \citenamefont
  {Ladoux}}]{saw2017topological}%
  \BibitemOpen
  \bibfield  {author} {\bibinfo {author} {\bibfnamefont {T.~B.}\ \bibnamefont
  {Saw}}, \bibinfo {author} {\bibfnamefont {A.}~\bibnamefont {Doostmohammadi}},
  \bibinfo {author} {\bibfnamefont {V.}~\bibnamefont {Nier}}, \bibinfo {author}
  {\bibfnamefont {L.}~\bibnamefont {Kocgozlu}}, \bibinfo {author}
  {\bibfnamefont {S.}~\bibnamefont {Thampi}}, \bibinfo {author} {\bibfnamefont
  {Y.}~\bibnamefont {Toyama}}, \bibinfo {author} {\bibfnamefont
  {P.}~\bibnamefont {Marcq}}, \bibinfo {author} {\bibfnamefont {C.~T.}\
  \bibnamefont {Lim}}, \bibinfo {author} {\bibfnamefont {J.~M.}\ \bibnamefont
  {Yeomans}}, \ and\ \bibinfo {author} {\bibfnamefont {B.}~\bibnamefont
  {Ladoux}},\ }\href@noop {} {\bibfield  {journal} {\bibinfo  {journal}
  {Nature}\ }\textbf {\bibinfo {volume} {544}},\ \bibinfo {pages} {212}
  (\bibinfo {year} {2017})}\BibitemShut {NoStop}%
\bibitem [{\citenamefont {Marchetti}\ \emph {et~al.}(2013)\citenamefont
  {Marchetti}, \citenamefont {Joanny}, \citenamefont {Ramaswamy}, \citenamefont
  {Liverpool}, \citenamefont {Prost}, \citenamefont {Rao},\ and\ \citenamefont
  {Simha}}]{marchetti2013hydrodynamics}%
  \BibitemOpen
  \bibfield  {author} {\bibinfo {author} {\bibfnamefont {M.~C.}\ \bibnamefont
  {Marchetti}}, \bibinfo {author} {\bibfnamefont {J.-F.}\ \bibnamefont
  {Joanny}}, \bibinfo {author} {\bibfnamefont {S.}~\bibnamefont {Ramaswamy}},
  \bibinfo {author} {\bibfnamefont {T.~B.}\ \bibnamefont {Liverpool}}, \bibinfo
  {author} {\bibfnamefont {J.}~\bibnamefont {Prost}}, \bibinfo {author}
  {\bibfnamefont {M.}~\bibnamefont {Rao}}, \ and\ \bibinfo {author}
  {\bibfnamefont {R.~A.}\ \bibnamefont {Simha}},\ }\href@noop {} {\bibfield
  {journal} {\bibinfo  {journal} {Reviews of Modern Physics}\ }\textbf
  {\bibinfo {volume} {85}},\ \bibinfo {pages} {1143} (\bibinfo {year}
  {2013})}\BibitemShut {NoStop}%
\bibitem [{\citenamefont {Peshkov}\ \emph {et~al.}(2014)\citenamefont
  {Peshkov}, \citenamefont {Bertin}, \citenamefont {Ginelli},\ and\
  \citenamefont {Chat{\'e}}}]{peshkov2014boltzmann}%
  \BibitemOpen
  \bibfield  {author} {\bibinfo {author} {\bibfnamefont {A.}~\bibnamefont
  {Peshkov}}, \bibinfo {author} {\bibfnamefont {E.}~\bibnamefont {Bertin}},
  \bibinfo {author} {\bibfnamefont {F.}~\bibnamefont {Ginelli}}, \ and\
  \bibinfo {author} {\bibfnamefont {H.}~\bibnamefont {Chat{\'e}}},\ }\href@noop
  {} {\bibfield  {journal} {\bibinfo  {journal} {The European Physical Journal
  Special Topics}\ }\textbf {\bibinfo {volume} {223}},\ \bibinfo {pages} {1315}
  (\bibinfo {year} {2014})}\BibitemShut {NoStop}%
\bibitem [{\citenamefont {Lee}\ and\ \citenamefont
  {Wolgemuth}(2011{\natexlab{a}})}]{lee2011crawling}%
  \BibitemOpen
  \bibfield  {author} {\bibinfo {author} {\bibfnamefont {P.}~\bibnamefont
  {Lee}}\ and\ \bibinfo {author} {\bibfnamefont {C.~W.}\ \bibnamefont
  {Wolgemuth}},\ }\href@noop {} {\bibfield  {journal} {\bibinfo  {journal}
  {PLoS Computational Biology}\ }\textbf {\bibinfo {volume} {7}},\ \bibinfo
  {pages} {e1002007} (\bibinfo {year} {2011}{\natexlab{a}})}\BibitemShut
  {NoStop}%
\bibitem [{\citenamefont {Lee}\ and\ \citenamefont
  {Wolgemuth}(2011{\natexlab{b}})}]{lee2011advent}%
  \BibitemOpen
  \bibfield  {author} {\bibinfo {author} {\bibfnamefont {P.}~\bibnamefont
  {Lee}}\ and\ \bibinfo {author} {\bibfnamefont {C.}~\bibnamefont
  {Wolgemuth}},\ }\href@noop {} {\bibfield  {journal} {\bibinfo  {journal}
  {Physical Review E}\ }\textbf {\bibinfo {volume} {83}},\ \bibinfo {pages}
  {061920} (\bibinfo {year} {2011}{\natexlab{b}})}\BibitemShut {NoStop}%
\bibitem [{\citenamefont {Gov}(2009)}]{gov2009traction}%
  \BibitemOpen
  \bibfield  {author} {\bibinfo {author} {\bibfnamefont {N.}~\bibnamefont
  {Gov}},\ }\href@noop {} {\bibfield  {journal} {\bibinfo  {journal} {HFSP
  Journal}\ }\textbf {\bibinfo {volume} {3}},\ \bibinfo {pages} {223} (\bibinfo
  {year} {2009})}\BibitemShut {NoStop}%
\bibitem [{\citenamefont {Camley}\ and\ \citenamefont
  {Rappel}(2017)}]{camley2017physical}%
  \BibitemOpen
  \bibfield  {author} {\bibinfo {author} {\bibfnamefont {B.~A.}\ \bibnamefont
  {Camley}}\ and\ \bibinfo {author} {\bibfnamefont {W.-J.}\ \bibnamefont
  {Rappel}},\ }\href@noop {} {\bibfield  {journal} {\bibinfo  {journal}
  {Journal of Physics D: Applied Physics}\ }\textbf {\bibinfo {volume} {50}},\
  \bibinfo {pages} {113002} (\bibinfo {year} {2017})}\BibitemShut {NoStop}%
\bibitem [{\citenamefont {Bi}\ \emph {et~al.}(2016)\citenamefont {Bi},
  \citenamefont {Yang}, \citenamefont {Marchetti},\ and\ \citenamefont
  {Manning}}]{bi2016motility}%
  \BibitemOpen
  \bibfield  {author} {\bibinfo {author} {\bibfnamefont {D.}~\bibnamefont
  {Bi}}, \bibinfo {author} {\bibfnamefont {X.}~\bibnamefont {Yang}}, \bibinfo
  {author} {\bibfnamefont {M.~C.}\ \bibnamefont {Marchetti}}, \ and\ \bibinfo
  {author} {\bibfnamefont {M.~L.}\ \bibnamefont {Manning}},\ }\href@noop {}
  {\bibfield  {journal} {\bibinfo  {journal} {Physical Review X}\ }\textbf
  {\bibinfo {volume} {6}},\ \bibinfo {pages} {021011} (\bibinfo {year}
  {2016})}\BibitemShut {NoStop}%
\bibitem [{\citenamefont {Henkes}\ \emph {et~al.}(2011)\citenamefont {Henkes},
  \citenamefont {Fily},\ and\ \citenamefont {Marchetti}}]{henkes2011active}%
  \BibitemOpen
  \bibfield  {author} {\bibinfo {author} {\bibfnamefont {S.}~\bibnamefont
  {Henkes}}, \bibinfo {author} {\bibfnamefont {Y.}~\bibnamefont {Fily}}, \ and\
  \bibinfo {author} {\bibfnamefont {M.~C.}\ \bibnamefont {Marchetti}},\
  }\href@noop {} {\bibfield  {journal} {\bibinfo  {journal} {Physical Review
  E}\ }\textbf {\bibinfo {volume} {84}},\ \bibinfo {pages} {040301} (\bibinfo
  {year} {2011})}\BibitemShut {NoStop}%
\bibitem [{\citenamefont {Etournay}\ \emph {et~al.}(2015)\citenamefont
  {Etournay}, \citenamefont {Popovi{\'c}}, \citenamefont {Merkel},
  \citenamefont {Nandi}, \citenamefont {Blasse}, \citenamefont {Aigouy},
  \citenamefont {Brandl}, \citenamefont {Myers}, \citenamefont {Salbreux},
  \citenamefont {J{\"u}licher} \emph {et~al.}}]{etournay2015interplay}%
  \BibitemOpen
  \bibfield  {author} {\bibinfo {author} {\bibfnamefont {R.}~\bibnamefont
  {Etournay}}, \bibinfo {author} {\bibfnamefont {M.}~\bibnamefont
  {Popovi{\'c}}}, \bibinfo {author} {\bibfnamefont {M.}~\bibnamefont {Merkel}},
  \bibinfo {author} {\bibfnamefont {A.}~\bibnamefont {Nandi}}, \bibinfo
  {author} {\bibfnamefont {C.}~\bibnamefont {Blasse}}, \bibinfo {author}
  {\bibfnamefont {B.}~\bibnamefont {Aigouy}}, \bibinfo {author} {\bibfnamefont
  {H.}~\bibnamefont {Brandl}}, \bibinfo {author} {\bibfnamefont
  {G.}~\bibnamefont {Myers}}, \bibinfo {author} {\bibfnamefont
  {G.}~\bibnamefont {Salbreux}}, \bibinfo {author} {\bibfnamefont
  {F.}~\bibnamefont {J{\"u}licher}},  \emph {et~al.},\ }\href@noop {}
  {\bibfield  {journal} {\bibinfo  {journal} {Elife}\ }\textbf {\bibinfo
  {volume} {4}},\ \bibinfo {pages} {e07090} (\bibinfo {year}
  {2015})}\BibitemShut {NoStop}%
\bibitem [{\citenamefont {Balasubramaniam}\ \emph {et~al.}(2021)\citenamefont
  {Balasubramaniam}, \citenamefont {Doostmohammadi}, \citenamefont {Saw},
  \citenamefont {Narayana}, \citenamefont {Mueller}, \citenamefont {Dang},
  \citenamefont {Thomas}, \citenamefont {Gupta}, \citenamefont {Sonam},
  \citenamefont {Yap} \emph {et~al.}}]{balasubramaniam2021investigating}%
  \BibitemOpen
  \bibfield  {author} {\bibinfo {author} {\bibfnamefont {L.}~\bibnamefont
  {Balasubramaniam}}, \bibinfo {author} {\bibfnamefont {A.}~\bibnamefont
  {Doostmohammadi}}, \bibinfo {author} {\bibfnamefont {T.~B.}\ \bibnamefont
  {Saw}}, \bibinfo {author} {\bibfnamefont {G.~H. N.~S.}\ \bibnamefont
  {Narayana}}, \bibinfo {author} {\bibfnamefont {R.}~\bibnamefont {Mueller}},
  \bibinfo {author} {\bibfnamefont {T.}~\bibnamefont {Dang}}, \bibinfo {author}
  {\bibfnamefont {M.}~\bibnamefont {Thomas}}, \bibinfo {author} {\bibfnamefont
  {S.}~\bibnamefont {Gupta}}, \bibinfo {author} {\bibfnamefont
  {S.}~\bibnamefont {Sonam}}, \bibinfo {author} {\bibfnamefont {A.~S.}\
  \bibnamefont {Yap}},  \emph {et~al.},\ }\href@noop {} {\bibfield  {journal}
  {\bibinfo  {journal} {Nature Materials}\ ,\ \bibinfo {pages} {1}} (\bibinfo
  {year} {2021})}\BibitemShut {NoStop}%
\bibitem [{\citenamefont {Barton}\ \emph {et~al.}(2017)\citenamefont {Barton},
  \citenamefont {Henkes}, \citenamefont {Weijer},\ and\ \citenamefont
  {Sknepnek}}]{barton2017active}%
  \BibitemOpen
  \bibfield  {author} {\bibinfo {author} {\bibfnamefont {D.~L.}\ \bibnamefont
  {Barton}}, \bibinfo {author} {\bibfnamefont {S.}~\bibnamefont {Henkes}},
  \bibinfo {author} {\bibfnamefont {C.~J.}\ \bibnamefont {Weijer}}, \ and\
  \bibinfo {author} {\bibfnamefont {R.}~\bibnamefont {Sknepnek}},\ }\href@noop
  {} {\bibfield  {journal} {\bibinfo  {journal} {PLoS Computational Biology}\
  }\textbf {\bibinfo {volume} {13}},\ \bibinfo {pages} {e1005569} (\bibinfo
  {year} {2017})}\BibitemShut {NoStop}%
\bibitem [{\citenamefont {Aranson}(2016)}]{aranson2016physical}%
  \BibitemOpen
  \bibfield  {author} {\bibinfo {author} {\bibfnamefont {I.~S.}\ \bibnamefont
  {Aranson}},\ }\href@noop {} {\emph {\bibinfo {title} {Physical Models of Cell
  Motility}}}\ (\bibinfo  {publisher} {Springer},\ \bibinfo {year}
  {2016})\BibitemShut {NoStop}%
\bibitem [{\citenamefont {Najem}\ and\ \citenamefont
  {Grant}(2016)}]{najem2016phase}%
  \BibitemOpen
  \bibfield  {author} {\bibinfo {author} {\bibfnamefont {S.}~\bibnamefont
  {Najem}}\ and\ \bibinfo {author} {\bibfnamefont {M.}~\bibnamefont {Grant}},\
  }\href@noop {} {\bibfield  {journal} {\bibinfo  {journal} {Physical Review
  E}\ }\textbf {\bibinfo {volume} {93}},\ \bibinfo {pages} {052405} (\bibinfo
  {year} {2016})}\BibitemShut {NoStop}%
\bibitem [{\citenamefont {Basan}\ \emph {et~al.}(2013)\citenamefont {Basan},
  \citenamefont {Elgeti}, \citenamefont {Hannezo}, \citenamefont {Rappel},\
  and\ \citenamefont {Levine}}]{basan2013alignment}%
  \BibitemOpen
  \bibfield  {author} {\bibinfo {author} {\bibfnamefont {M.}~\bibnamefont
  {Basan}}, \bibinfo {author} {\bibfnamefont {J.}~\bibnamefont {Elgeti}},
  \bibinfo {author} {\bibfnamefont {E.}~\bibnamefont {Hannezo}}, \bibinfo
  {author} {\bibfnamefont {W.-J.}\ \bibnamefont {Rappel}}, \ and\ \bibinfo
  {author} {\bibfnamefont {H.}~\bibnamefont {Levine}},\ }\href@noop {}
  {\bibfield  {journal} {\bibinfo  {journal} {Proceedings of the National
  Academy of Sciences}\ }\textbf {\bibinfo {volume} {110}},\ \bibinfo {pages}
  {2452} (\bibinfo {year} {2013})}\BibitemShut {NoStop}%
\bibitem [{\citenamefont {Sep{\'u}lveda}\ \emph {et~al.}(2013)\citenamefont
  {Sep{\'u}lveda}, \citenamefont {Petitjean}, \citenamefont {Cochet},
  \citenamefont {Grasland-Mongrain}, \citenamefont {Silberzan},\ and\
  \citenamefont {Hakim}}]{sepulveda2013collective}%
  \BibitemOpen
  \bibfield  {author} {\bibinfo {author} {\bibfnamefont {N.}~\bibnamefont
  {Sep{\'u}lveda}}, \bibinfo {author} {\bibfnamefont {L.}~\bibnamefont
  {Petitjean}}, \bibinfo {author} {\bibfnamefont {O.}~\bibnamefont {Cochet}},
  \bibinfo {author} {\bibfnamefont {E.}~\bibnamefont {Grasland-Mongrain}},
  \bibinfo {author} {\bibfnamefont {P.}~\bibnamefont {Silberzan}}, \ and\
  \bibinfo {author} {\bibfnamefont {V.}~\bibnamefont {Hakim}},\ }\href@noop {}
  {\bibfield  {journal} {\bibinfo  {journal} {PLoS Computational Biology}\
  }\textbf {\bibinfo {volume} {9}},\ \bibinfo {pages} {e1002944} (\bibinfo
  {year} {2013})}\BibitemShut {NoStop}%
\bibitem [{\citenamefont {Monfared}\ \emph {et~al.}(2021)\citenamefont
  {Monfared}, \citenamefont {Ravichandran}, \citenamefont {Andrade},\ and\
  \citenamefont {Doostmohammadi}}]{monfared2021mechanics}%
  \BibitemOpen
  \bibfield  {author} {\bibinfo {author} {\bibfnamefont {S.}~\bibnamefont
  {Monfared}}, \bibinfo {author} {\bibfnamefont {G.}~\bibnamefont
  {Ravichandran}}, \bibinfo {author} {\bibfnamefont {J.~E.}\ \bibnamefont
  {Andrade}}, \ and\ \bibinfo {author} {\bibfnamefont {A.}~\bibnamefont
  {Doostmohammadi}},\ }\href@noop {} {\bibfield  {journal} {\bibinfo  {journal}
  {arXiv preprint arXiv:2108.07657}\ } (\bibinfo {year} {2021})}\BibitemShut
  {NoStop}%
\bibitem [{\citenamefont {Mueller}\ \emph {et~al.}(2019)\citenamefont
  {Mueller}, \citenamefont {Yeomans},\ and\ \citenamefont
  {Doostmohammadi}}]{mueller2019emergence}%
  \BibitemOpen
  \bibfield  {author} {\bibinfo {author} {\bibfnamefont {R.}~\bibnamefont
  {Mueller}}, \bibinfo {author} {\bibfnamefont {J.~M.}\ \bibnamefont
  {Yeomans}}, \ and\ \bibinfo {author} {\bibfnamefont {A.}~\bibnamefont
  {Doostmohammadi}},\ }\href@noop {} {\bibfield  {journal} {\bibinfo  {journal}
  {Physical Review Letters}\ }\textbf {\bibinfo {volume} {122}},\ \bibinfo
  {pages} {048004} (\bibinfo {year} {2019})}\BibitemShut {NoStop}%
\bibitem [{\citenamefont {Zhang}\ \emph {et~al.}(2020)\citenamefont {Zhang},
  \citenamefont {Mueller}, \citenamefont {Doostmohammadi},\ and\ \citenamefont
  {Yeomans}}]{zhang2020active}%
  \BibitemOpen
  \bibfield  {author} {\bibinfo {author} {\bibfnamefont {G.}~\bibnamefont
  {Zhang}}, \bibinfo {author} {\bibfnamefont {R.}~\bibnamefont {Mueller}},
  \bibinfo {author} {\bibfnamefont {A.}~\bibnamefont {Doostmohammadi}}, \ and\
  \bibinfo {author} {\bibfnamefont {J.~M.}\ \bibnamefont {Yeomans}},\
  }\href@noop {} {\bibfield  {journal} {\bibinfo  {journal} {Journal of the
  Royal Society Interface}\ }\textbf {\bibinfo {volume} {17}},\ \bibinfo
  {pages} {20200312} (\bibinfo {year} {2020})}\BibitemShut {NoStop}%
\bibitem [{\citenamefont {Zhang}\ and\ \citenamefont
  {Yeomans}(2021)}]{zhang2021active}%
  \BibitemOpen
  \bibfield  {author} {\bibinfo {author} {\bibfnamefont {G.}~\bibnamefont
  {Zhang}}\ and\ \bibinfo {author} {\bibfnamefont {J.~M.}\ \bibnamefont
  {Yeomans}},\ }\href@noop {} {\bibfield  {journal} {\bibinfo  {journal} {arXiv
  preprint arXiv:2111.14401}\ } (\bibinfo {year} {2021})}\BibitemShut {NoStop}%
\bibitem [{\citenamefont {Palmieri}\ \emph {et~al.}(2015)\citenamefont
  {Palmieri}, \citenamefont {Bresler}, \citenamefont {Wirtz},\ and\
  \citenamefont {Grant}}]{palmieri2015multiple}%
  \BibitemOpen
  \bibfield  {author} {\bibinfo {author} {\bibfnamefont {B.}~\bibnamefont
  {Palmieri}}, \bibinfo {author} {\bibfnamefont {Y.}~\bibnamefont {Bresler}},
  \bibinfo {author} {\bibfnamefont {D.}~\bibnamefont {Wirtz}}, \ and\ \bibinfo
  {author} {\bibfnamefont {M.}~\bibnamefont {Grant}},\ }\href@noop {}
  {\bibfield  {journal} {\bibinfo  {journal} {Scientific Reports}\ }\textbf
  {\bibinfo {volume} {5}},\ \bibinfo {pages} {1} (\bibinfo {year}
  {2015})}\BibitemShut {NoStop}%
\bibitem [{\citenamefont {Helfrich}(1973)}]{PMID:4273690}%
  \BibitemOpen
  \bibfield  {author} {\bibinfo {author} {\bibfnamefont {W.}~\bibnamefont
  {Helfrich}},\ }\href@noop {} {\bibfield  {journal} {\bibinfo  {journal} {Z
  Naturforsch C}\ }\textbf {\bibinfo {volume} {28}},\ \bibinfo {pages} {693}
  (\bibinfo {year} {1973})}\BibitemShut {NoStop}%
\bibitem [{\citenamefont {Shao}\ \emph {et~al.}(2010)\citenamefont {Shao},
  \citenamefont {Rappel},\ and\ \citenamefont
  {Levine}}]{shao2010computational}%
  \BibitemOpen
  \bibfield  {author} {\bibinfo {author} {\bibfnamefont {D.}~\bibnamefont
  {Shao}}, \bibinfo {author} {\bibfnamefont {W.-J.}\ \bibnamefont {Rappel}}, \
  and\ \bibinfo {author} {\bibfnamefont {H.}~\bibnamefont {Levine}},\
  }\href@noop {} {\bibfield  {journal} {\bibinfo  {journal} {Physical Review
  Letters}\ }\textbf {\bibinfo {volume} {105}},\ \bibinfo {pages} {108104}
  (\bibinfo {year} {2010})}\BibitemShut {NoStop}%
\bibitem [{\citenamefont {Biben}\ and\ \citenamefont
  {Misbah}(2003)}]{biben2003tumbling}%
  \BibitemOpen
  \bibfield  {author} {\bibinfo {author} {\bibfnamefont {T.}~\bibnamefont
  {Biben}}\ and\ \bibinfo {author} {\bibfnamefont {C.}~\bibnamefont {Misbah}},\
  }\href@noop {} {\bibfield  {journal} {\bibinfo  {journal} {Physical Review
  E}\ }\textbf {\bibinfo {volume} {67}},\ \bibinfo {pages} {031908} (\bibinfo
  {year} {2003})}\BibitemShut {NoStop}%
\bibitem [{\citenamefont {Peyret}\ \emph {et~al.}(2019)\citenamefont {Peyret},
  \citenamefont {Mueller}, \citenamefont {d’Alessandro}, \citenamefont
  {Begnaud}, \citenamefont {Marcq}, \citenamefont {M{\`e}ge}, \citenamefont
  {Yeomans}, \citenamefont {Doostmohammadi},\ and\ \citenamefont
  {Ladoux}}]{peyret2019sustained}%
  \BibitemOpen
  \bibfield  {author} {\bibinfo {author} {\bibfnamefont {G.}~\bibnamefont
  {Peyret}}, \bibinfo {author} {\bibfnamefont {R.}~\bibnamefont {Mueller}},
  \bibinfo {author} {\bibfnamefont {J.}~\bibnamefont {d’Alessandro}},
  \bibinfo {author} {\bibfnamefont {S.}~\bibnamefont {Begnaud}}, \bibinfo
  {author} {\bibfnamefont {P.}~\bibnamefont {Marcq}}, \bibinfo {author}
  {\bibfnamefont {R.-M.}\ \bibnamefont {M{\`e}ge}}, \bibinfo {author}
  {\bibfnamefont {J.~M.}\ \bibnamefont {Yeomans}}, \bibinfo {author}
  {\bibfnamefont {A.}~\bibnamefont {Doostmohammadi}}, \ and\ \bibinfo {author}
  {\bibfnamefont {B.}~\bibnamefont {Ladoux}},\ }\href@noop {} {\bibfield
  {journal} {\bibinfo  {journal} {Biophysical Journal}\ }\textbf {\bibinfo
  {volume} {117}},\ \bibinfo {pages} {464} (\bibinfo {year}
  {2019})}\BibitemShut {NoStop}%
\bibitem [{\citenamefont {Simha}\ and\ \citenamefont
  {Ramaswamy}(2002)}]{simha2002hydrodynamic}%
  \BibitemOpen
  \bibfield  {author} {\bibinfo {author} {\bibfnamefont {R.~A.}\ \bibnamefont
  {Simha}}\ and\ \bibinfo {author} {\bibfnamefont {S.}~\bibnamefont
  {Ramaswamy}},\ }\href@noop {} {\bibfield  {journal} {\bibinfo  {journal}
  {Physical Review Letters}\ }\textbf {\bibinfo {volume} {89}},\ \bibinfo
  {pages} {058101} (\bibinfo {year} {2002})}\BibitemShut {NoStop}%
\bibitem [{\citenamefont {Ramaswamy}\ and\ \citenamefont
  {Rao}(2007)}]{ramaswamy2007active}%
  \BibitemOpen
  \bibfield  {author} {\bibinfo {author} {\bibfnamefont {S.}~\bibnamefont
  {Ramaswamy}}\ and\ \bibinfo {author} {\bibfnamefont {M.}~\bibnamefont
  {Rao}},\ }\href@noop {} {\bibfield  {journal} {\bibinfo  {journal} {New
  Journal of Physics}\ }\textbf {\bibinfo {volume} {9}},\ \bibinfo {pages}
  {423} (\bibinfo {year} {2007})}\BibitemShut {NoStop}%
\bibitem [{\citenamefont {Thampi}\ \emph {et~al.}(2014)\citenamefont {Thampi},
  \citenamefont {Golestanian},\ and\ \citenamefont
  {Yeomans}}]{thampi2014instabilities}%
  \BibitemOpen
  \bibfield  {author} {\bibinfo {author} {\bibfnamefont {S.~P.}\ \bibnamefont
  {Thampi}}, \bibinfo {author} {\bibfnamefont {R.}~\bibnamefont {Golestanian}},
  \ and\ \bibinfo {author} {\bibfnamefont {J.~M.}\ \bibnamefont {Yeomans}},\
  }\href@noop {} {\bibfield  {journal} {\bibinfo  {journal} {EPL (Europhysics
  Letters)}\ }\textbf {\bibinfo {volume} {105}},\ \bibinfo {pages} {18001}
  (\bibinfo {year} {2014})}\BibitemShut {NoStop}%
\bibitem [{\citenamefont {Giomi}\ \emph {et~al.}(2011)\citenamefont {Giomi},
  \citenamefont {Mahadevan}, \citenamefont {Chakraborty},\ and\ \citenamefont
  {Hagan}}]{giomi2011excitable}%
  \BibitemOpen
  \bibfield  {author} {\bibinfo {author} {\bibfnamefont {L.}~\bibnamefont
  {Giomi}}, \bibinfo {author} {\bibfnamefont {L.}~\bibnamefont {Mahadevan}},
  \bibinfo {author} {\bibfnamefont {B.}~\bibnamefont {Chakraborty}}, \ and\
  \bibinfo {author} {\bibfnamefont {M.}~\bibnamefont {Hagan}},\ }\href@noop {}
  {\bibfield  {journal} {\bibinfo  {journal} {Physical Review Letters}\
  }\textbf {\bibinfo {volume} {106}},\ \bibinfo {pages} {218101} (\bibinfo
  {year} {2011})}\BibitemShut {NoStop}%
\bibitem [{\citenamefont {Giomi}\ \emph {et~al.}(2013)\citenamefont {Giomi},
  \citenamefont {Bowick}, \citenamefont {Ma},\ and\ \citenamefont
  {Marchetti}}]{giomi2013defect}%
  \BibitemOpen
  \bibfield  {author} {\bibinfo {author} {\bibfnamefont {L.}~\bibnamefont
  {Giomi}}, \bibinfo {author} {\bibfnamefont {M.~J.}\ \bibnamefont {Bowick}},
  \bibinfo {author} {\bibfnamefont {X.}~\bibnamefont {Ma}}, \ and\ \bibinfo
  {author} {\bibfnamefont {M.~C.}\ \bibnamefont {Marchetti}},\ }\href@noop {}
  {\bibfield  {journal} {\bibinfo  {journal} {Physical Review Letters}\
  }\textbf {\bibinfo {volume} {110}},\ \bibinfo {pages} {228101} (\bibinfo
  {year} {2013})}\BibitemShut {NoStop}%
\bibitem [{\citenamefont {Meacock}\ \emph {et~al.}(2021)\citenamefont
  {Meacock}, \citenamefont {Doostmohammadi}, \citenamefont {Foster},
  \citenamefont {Yeomans},\ and\ \citenamefont {Durham}}]{meacock2021bacteria}%
  \BibitemOpen
  \bibfield  {author} {\bibinfo {author} {\bibfnamefont {O.~J.}\ \bibnamefont
  {Meacock}}, \bibinfo {author} {\bibfnamefont {A.}~\bibnamefont
  {Doostmohammadi}}, \bibinfo {author} {\bibfnamefont {K.~R.}\ \bibnamefont
  {Foster}}, \bibinfo {author} {\bibfnamefont {J.~M.}\ \bibnamefont {Yeomans}},
  \ and\ \bibinfo {author} {\bibfnamefont {W.~M.}\ \bibnamefont {Durham}},\
  }\href@noop {} {\bibfield  {journal} {\bibinfo  {journal} {Nature Physics}\
  }\textbf {\bibinfo {volume} {17}},\ \bibinfo {pages} {205} (\bibinfo {year}
  {2021})}\BibitemShut {NoStop}%
\bibitem [{\citenamefont {Sanchez}\ \emph {et~al.}(2012)\citenamefont
  {Sanchez}, \citenamefont {Chen}, \citenamefont {DeCamp}, \citenamefont
  {Heymann},\ and\ \citenamefont {Dogic}}]{Sanchez2012}%
  \BibitemOpen
  \bibfield  {author} {\bibinfo {author} {\bibfnamefont {T.}~\bibnamefont
  {Sanchez}}, \bibinfo {author} {\bibfnamefont {D.~T.~N.}\ \bibnamefont
  {Chen}}, \bibinfo {author} {\bibfnamefont {S.~J.}\ \bibnamefont {DeCamp}},
  \bibinfo {author} {\bibfnamefont {M.}~\bibnamefont {Heymann}}, \ and\
  \bibinfo {author} {\bibfnamefont {Z.}~\bibnamefont {Dogic}},\ }\href@noop {}
  {\bibfield  {journal} {\bibinfo  {journal} {Nature}\ }\textbf {\bibinfo
  {volume} {491}},\ \bibinfo {pages} {431} (\bibinfo {year}
  {2012})}\BibitemShut {NoStop}%
\bibitem [{\citenamefont {Kumar}\ \emph {et~al.}(2018)\citenamefont {Kumar},
  \citenamefont {Zhang}, \citenamefont {De~Pablo},\ and\ \citenamefont
  {Gardel}}]{kumar2018tunable}%
  \BibitemOpen
  \bibfield  {author} {\bibinfo {author} {\bibfnamefont {N.}~\bibnamefont
  {Kumar}}, \bibinfo {author} {\bibfnamefont {R.}~\bibnamefont {Zhang}},
  \bibinfo {author} {\bibfnamefont {J.~J.}\ \bibnamefont {De~Pablo}}, \ and\
  \bibinfo {author} {\bibfnamefont {M.~L.}\ \bibnamefont {Gardel}},\
  }\href@noop {} {\bibfield  {journal} {\bibinfo  {journal} {Science Advances}\
  }\textbf {\bibinfo {volume} {4}},\ \bibinfo {pages} {eaat7779} (\bibinfo
  {year} {2018})}\BibitemShut {NoStop}%
\bibitem [{\citenamefont {Wenzel}\ and\ \citenamefont
  {Voigt}(2021)}]{wenzel2021multiphase}%
  \BibitemOpen
  \bibfield  {author} {\bibinfo {author} {\bibfnamefont {D.}~\bibnamefont
  {Wenzel}}\ and\ \bibinfo {author} {\bibfnamefont {A.}~\bibnamefont {Voigt}},\
  }\href@noop {} {\bibfield  {journal} {\bibinfo  {journal} {arXiv preprint
  arXiv:2106.10552}\ } (\bibinfo {year} {2021})}\BibitemShut {NoStop}%
\bibitem [{\citenamefont {Giomi}(2015)}]{giomi2015geometry}%
  \BibitemOpen
  \bibfield  {author} {\bibinfo {author} {\bibfnamefont {L.}~\bibnamefont
  {Giomi}},\ }\href@noop {} {\bibfield  {journal} {\bibinfo  {journal}
  {Physical Review X}\ }\textbf {\bibinfo {volume} {5}},\ \bibinfo {pages}
  {031003} (\bibinfo {year} {2015})}\BibitemShut {NoStop}%
\bibitem [{\citenamefont {Mart{\'\i}nez-Prat}\ \emph
  {et~al.}(2019)\citenamefont {Mart{\'\i}nez-Prat}, \citenamefont
  {Ign{\'e}s-Mullol}, \citenamefont {Casademunt},\ and\ \citenamefont
  {Sagu{\'e}s}}]{martinez2019selection}%
  \BibitemOpen
  \bibfield  {author} {\bibinfo {author} {\bibfnamefont {B.}~\bibnamefont
  {Mart{\'\i}nez-Prat}}, \bibinfo {author} {\bibfnamefont {J.}~\bibnamefont
  {Ign{\'e}s-Mullol}}, \bibinfo {author} {\bibfnamefont {J.}~\bibnamefont
  {Casademunt}}, \ and\ \bibinfo {author} {\bibfnamefont {F.}~\bibnamefont
  {Sagu{\'e}s}},\ }\href@noop {} {\bibfield  {journal} {\bibinfo  {journal}
  {Nature Physics}\ }\textbf {\bibinfo {volume} {15}},\ \bibinfo {pages} {362}
  (\bibinfo {year} {2019})}\BibitemShut {NoStop}%
\bibitem [{\citenamefont {P{\'e}rez-Gonz{\'a}lez}\ \emph
  {et~al.}(2019)\citenamefont {P{\'e}rez-Gonz{\'a}lez}, \citenamefont {Alert},
  \citenamefont {Blanch-Mercader}, \citenamefont {G{\'o}mez-Gonz{\'a}lez},
  \citenamefont {Kolodziej}, \citenamefont {Bazellieres}, \citenamefont
  {Casademunt},\ and\ \citenamefont {Trepat}}]{perez2019active}%
  \BibitemOpen
  \bibfield  {author} {\bibinfo {author} {\bibfnamefont {C.}~\bibnamefont
  {P{\'e}rez-Gonz{\'a}lez}}, \bibinfo {author} {\bibfnamefont {R.}~\bibnamefont
  {Alert}}, \bibinfo {author} {\bibfnamefont {C.}~\bibnamefont
  {Blanch-Mercader}}, \bibinfo {author} {\bibfnamefont {M.}~\bibnamefont
  {G{\'o}mez-Gonz{\'a}lez}}, \bibinfo {author} {\bibfnamefont {T.}~\bibnamefont
  {Kolodziej}}, \bibinfo {author} {\bibfnamefont {E.}~\bibnamefont
  {Bazellieres}}, \bibinfo {author} {\bibfnamefont {J.}~\bibnamefont
  {Casademunt}}, \ and\ \bibinfo {author} {\bibfnamefont {X.}~\bibnamefont
  {Trepat}},\ }\href@noop {} {\bibfield  {journal} {\bibinfo  {journal} {Nature
  Physics}\ }\textbf {\bibinfo {volume} {15}},\ \bibinfo {pages} {79} (\bibinfo
  {year} {2019})}\BibitemShut {NoStop}%
\bibitem [{\citenamefont {Alert}\ \emph {et~al.}(2019)\citenamefont {Alert},
  \citenamefont {Blanch-Mercader},\ and\ \citenamefont
  {Casademunt}}]{alert2019active}%
  \BibitemOpen
  \bibfield  {author} {\bibinfo {author} {\bibfnamefont {R.}~\bibnamefont
  {Alert}}, \bibinfo {author} {\bibfnamefont {C.}~\bibnamefont
  {Blanch-Mercader}}, \ and\ \bibinfo {author} {\bibfnamefont {J.}~\bibnamefont
  {Casademunt}},\ }\href@noop {} {\bibfield  {journal} {\bibinfo  {journal}
  {Physical Review Letters}\ }\textbf {\bibinfo {volume} {122}},\ \bibinfo
  {pages} {088104} (\bibinfo {year} {2019})}\BibitemShut {NoStop}%
\bibitem [{\citenamefont {Kim}\ \emph {et~al.}(2021)\citenamefont {Kim},
  \citenamefont {Pochitaloff}, \citenamefont {Stooke-Vaughan},\ and\
  \citenamefont {Camp{\`a}s}}]{kim2021embryonic}%
  \BibitemOpen
  \bibfield  {author} {\bibinfo {author} {\bibfnamefont {S.}~\bibnamefont
  {Kim}}, \bibinfo {author} {\bibfnamefont {M.}~\bibnamefont {Pochitaloff}},
  \bibinfo {author} {\bibfnamefont {G.~A.}\ \bibnamefont {Stooke-Vaughan}}, \
  and\ \bibinfo {author} {\bibfnamefont {O.}~\bibnamefont {Camp{\`a}s}},\
  }\href@noop {} {\bibfield  {journal} {\bibinfo  {journal} {Nature Physics}\
  ,\ \bibinfo {pages} {1}} (\bibinfo {year} {2021})}\BibitemShut {NoStop}%
\bibitem [{\citenamefont {Zheng}\ \emph {et~al.}(2017)\citenamefont {Zheng},
  \citenamefont {Han}, \citenamefont {Chiu}, \citenamefont {Yip}, \citenamefont
  {Boichat}, \citenamefont {Zhu}, \citenamefont {Zhong},\ and\ \citenamefont
  {Matsudaira}}]{zheng2017epithelial}%
  \BibitemOpen
  \bibfield  {author} {\bibinfo {author} {\bibfnamefont {J.~Y.}\ \bibnamefont
  {Zheng}}, \bibinfo {author} {\bibfnamefont {S.~P.}\ \bibnamefont {Han}},
  \bibinfo {author} {\bibfnamefont {Y.-J.}\ \bibnamefont {Chiu}}, \bibinfo
  {author} {\bibfnamefont {A.~K.}\ \bibnamefont {Yip}}, \bibinfo {author}
  {\bibfnamefont {N.}~\bibnamefont {Boichat}}, \bibinfo {author} {\bibfnamefont
  {S.~W.}\ \bibnamefont {Zhu}}, \bibinfo {author} {\bibfnamefont
  {J.}~\bibnamefont {Zhong}}, \ and\ \bibinfo {author} {\bibfnamefont
  {P.}~\bibnamefont {Matsudaira}},\ }\href@noop {} {\bibfield  {journal}
  {\bibinfo  {journal} {Biophysical Journal}\ }\textbf {\bibinfo {volume}
  {113}},\ \bibinfo {pages} {1585} (\bibinfo {year} {2017})}\BibitemShut
  {NoStop}%
\bibitem [{\citenamefont {Xi}\ \emph {et~al.}(2019)\citenamefont {Xi},
  \citenamefont {Saw}, \citenamefont {Delacour}, \citenamefont {Lim},\ and\
  \citenamefont {Ladoux}}]{xi2019material}%
  \BibitemOpen
  \bibfield  {author} {\bibinfo {author} {\bibfnamefont {W.}~\bibnamefont
  {Xi}}, \bibinfo {author} {\bibfnamefont {T.~B.}\ \bibnamefont {Saw}},
  \bibinfo {author} {\bibfnamefont {D.}~\bibnamefont {Delacour}}, \bibinfo
  {author} {\bibfnamefont {C.~T.}\ \bibnamefont {Lim}}, \ and\ \bibinfo
  {author} {\bibfnamefont {B.}~\bibnamefont {Ladoux}},\ }\href@noop {}
  {\bibfield  {journal} {\bibinfo  {journal} {Nature Reviews Materials}\
  }\textbf {\bibinfo {volume} {4}},\ \bibinfo {pages} {23} (\bibinfo {year}
  {2019})}\BibitemShut {NoStop}%
\bibitem [{\citenamefont {Sonam}\ \emph {et~al.}(2022)\citenamefont {Sonam},
  \citenamefont {Balasubramaniam}, \citenamefont {Lin}, \citenamefont {Ivan},
  \citenamefont {Jaum{\`a}}, \citenamefont {Jebane}, \citenamefont {Karnat},
  \citenamefont {Toyama}, \citenamefont {Marcq}, \citenamefont {Prost} \emph
  {et~al.}}]{sonam2022mechanical}%
  \BibitemOpen
  \bibfield  {author} {\bibinfo {author} {\bibfnamefont {S.}~\bibnamefont
  {Sonam}}, \bibinfo {author} {\bibfnamefont {L.}~\bibnamefont
  {Balasubramaniam}}, \bibinfo {author} {\bibfnamefont {S.-Z.}\ \bibnamefont
  {Lin}}, \bibinfo {author} {\bibfnamefont {Y.~M.~Y.}\ \bibnamefont {Ivan}},
  \bibinfo {author} {\bibfnamefont {I.~P.}\ \bibnamefont {Jaum{\`a}}}, \bibinfo
  {author} {\bibfnamefont {C.}~\bibnamefont {Jebane}}, \bibinfo {author}
  {\bibfnamefont {M.}~\bibnamefont {Karnat}}, \bibinfo {author} {\bibfnamefont
  {Y.}~\bibnamefont {Toyama}}, \bibinfo {author} {\bibfnamefont
  {P.}~\bibnamefont {Marcq}}, \bibinfo {author} {\bibfnamefont
  {J.}~\bibnamefont {Prost}},  \emph {et~al.},\ }\href@noop {} {\bibfield
  {journal} {\bibinfo  {journal} {bioRxiv}\ } (\bibinfo {year}
  {2022})}\BibitemShut {NoStop}%
\end{thebibliography}%


%

\end{document}